\documentclass[preprint,12pt,3p]{article}
\usepackage{jinstpub}
\usepackage{graphicx}
\usepackage{caption}
\usepackage{subcaption}
\usepackage{dcolumn}
\usepackage{bm}
\usepackage{xcolor}
\usepackage[normalem]{ulem} 
\usepackage{hyperref}
\usepackage[switch,mathlines]{lineno}

\usepackage{longtable}
\makeatletter
\gdef\@fpheader{}
\makeatother
\title{Comparative scintillation performance of EJ-309, EJ-276, and a novel organic glass}

\author[a,1]{T.~A.~Laplace,\note{Corresponding author.}}
\author[a,b,1]{B.~L.~Goldblum,}
\author[a,2]{J.~E.~Bevins,\note{Now at Department of Engineering Physics, Air Force Institute of Technology, Wright-Patterson Air Force Base, OH 45433 USA.}}
\author[c]{D.~L.~Bleuel,}
\author[b]{E.~Bourret,}
\author[a]{J.~A.~Brown,}
\author[d]{E.~J.~Callaghan,}
\author[e]{J.~S.~Carlson,}
\author[e]{P.~L.~Feng,}
\author[a]{G.~Gabella,}
\author[a,3]{K.~P.~Harrig,\note{Now at Department of Physics, University of California, Davis, CA 95616 USA.}}
\author[a]{J.~J.~Manfredi,}
\author[a]{C.~Moore,}
\author[b]{F.~Moretti,}
\author[a]{M.~Shinner,}
\author[a]{A.~Sweet,}
\author[a,3]{Z.~W.~Sweger}

\affiliation[a]{Department of Nuclear Engineering, University of California,\\ Berkeley, CA 94720 USA.}
\affiliation[b]{Lawrence Berkeley National Laboratory,\\ Berkeley, CA, 94720 USA.}
\affiliation[c]{Lawrence Livermore National Laboratory,\\ Livermore, CA, 94550 USA.}
\affiliation[d]{Department of Physics, University of California,\\ Berkeley, CA, 94720 USA.}
\affiliation[e]{Sandia National Laboratories,\\ Livermore, CA, 94550 USA.}

\emailAdd{lapthi@berkeley.edu; bethany@nuc.berkeley.edu}

\abstract{An organic glass scintillator developed by Sandia National Laboratories was characterized in terms of its light output and pulse shape discrimination (PSD) properties and compared to commercial liquid (EJ-309) and plastic (EJ-276) organic scintillators. The electron light output was determined through relative comparison of the $^{137}$Cs Compton edge location. The proton light yield was measured using a double time-of-flight technique at the 88-Inch Cyclotron at Lawrence Berkeley National Laboratory. Using a tunable broad-spectrum neutron source and an array of pulse-shape-discriminating observation scintillators, a continuous measurement of the proton light yield was performed for EJ-309 (200~keV$-$3.2~MeV), EJ-276 (170~keV$-$4.9~MeV), and the organic glass (50~keV$-$20~MeV). Finally, the PSD properties of the organic glass, EJ-309, and EJ-276 were evaluated using an AmBe source and compared via a figure-of-merit metric. The organic glass exhibited a higher electron light output than both EJ-309 and EJ-276. Its proton light yield and PSD performance were comparable to EJ-309 and superior to that of EJ-276. With these performance characteristics, the organic glass scintillator is well poised to replace current state-of-the-art PSD-capable scintillators in a range of fast neutron detection applications.}


\begin{document}
	\maketitle
	\flushbottom
	
\section{Introduction}

Pulse shape discrimination (PSD) in organic scintillators---the ability to distinguish the type of incident ionizing radiation based on differences in the temporal response of the scintillation light---has wide ranging applications in basic and applied nuclear physics, homeland security, proliferation detection, and treaty verification~\cite{Akerib2018,Ellis2016,Flaska2007,Hamel2017}. Historically, melt-grown trans-stilbene crystals (and more recently solution-grown trans-stilbene crystals and organic liquids such as EJ-309) have been the scintillators of choice when PSD properties are required~\cite{zaitseva2015}. While stilbene demonstrates superior light output, challenges are introduced in its application due to the fragility of the material, owing to its monoclinic crystal structure, and anisotropy of the scintillation response~\cite{reddy2006, weldon2018}. Stilbene is also expensive and crystals are difficult to grow in large scale, despite recent advancements in solution growth methods~\cite{zaitseva2011}. Liquid organic scintillators are appealing alternatives to organic crystals as they are relatively inexpensive and can be easily prepared in varied shape and volume. Eljen Technology's EJ-309, an organic liquid scintillator with high light output, high flash point, and relatively low chemical toxicity demonstrates excellent PSD performance~\cite{Kaplan2013}. However, liquids present safety and electrical hazards from potential leaks that may hinder their use in certain applications.

In contrast, plastic scintillators are robust, easily machinable, and capable of operation in harsh environments. They are inexpensive and can be produced in large quantity making them attractive for a wide range of applications. While PSD in plastic scintillators was initially demonstrated by Brooks et al.\ in 1960~\cite{brooks1960}, it wasn't until recent years that PSD-capable plastic organic scintillators were developed in a manner suitable for commercialization~\cite{zaitseva2012}. Eljen Technology offered two developmental PSD plastic formulations, EJ-299-33A and EJ-299-34~\cite{nyibule2013,pozzi2013,Payne2015}, and recently introduced EJ-276 as an optimized composition reputed to demonstrate PSD performance competitive with EJ-309~\cite{zaitseva2018,taggart2018}. Despite these advances, the scintillation efficiency of EJ-276 at 1~MeVee is $\sim$70\% that of EJ-309~\cite{276spec,309spec}, reducing the efficacy of this PSD-capable plastic scintillator in certain applications.

In 2016, Carlson et al.\ introduced a new class of organic scintillator---a PSD-capable organic glass~\cite{CARLSON2016_OGlassNIM,Carlson2017_OGlassJACS}. The molecules in an organic glass are arranged in random orientations, leading to isotropic light output and the absence of defined cleavage planes, with a resulting structure more resistant to failure due to fracture relative to crystals. The organic glass is composed of inexpensive materials and the simplicity of the melt-casting process, with its short duration and wide tolerances, results in low manufacturing costs. The light yield of the organic glass was shown to be higher than solution-grown trans-stilbene with PSD performance that near rivals this material~\cite{CARLSON2016_OGlassNIM}. With such properties, this material is a potential replacement for PSD-capable liquid and plastic organic scintillators in some applications. To evaluate this proposition, this work provides a quantitative comparison of the relative electron light yield, proton light yield, and PSD properties of EJ-309, EJ-276, and the organic glass. 


\section{Organic glass composition}
\label{o-glass}

The organic glass composition studied in this work was a 90:10 mixture of bis(9,9-dimethyl-9H-fluoren-2-yl)(dimethyl)silane (`P2') : tris(9,9-dimethyl-9H-fluoren-2-yl)(methyl)silane (`P3'), along with 0.2 wt. \% of 1,4-Bis(2-methylstyryl)benzene (`bis-MSB'). The scintillator monolith was melt-cast in December 2018 in an aluminum mold according to procedures described in the literature \cite{Carlson2017_OGlassJACS}. Briefly, this involved heating a mixture of P2, P3, and bis-MSB above 140$^{\circ}$C and pouring the molten mixture into a pre-heated aluminum split mold. The material was allowed to cool to room temperature in a temperature-controlled oven for a period of several hours.


\section{Materials, data acquisition, and signal processing}
\label{materials}

Each of the scintillating media examined in this work were configured as 5.08 cm dia.\ $\times$ 5.08 cm length right circular cylinders directly coupled to a photomultiplier tube (PMT). The EJ-309 liquid was encapsulated within an aluminum cell internally coated with EJ-520 titanium dioxide reflective paint~\cite{EJ520spec}. Optical coupling was accomplished using a thin layer of EJ-550 optical grade silicone grease directly to the PMT borosilicate glass window for EJ-276 and via a 6.3-mm-thick Pyrex window for the EJ-309 cell. Because the organic glass is partially soluble in silicone grease, a polyvinyl alcohol solution was used for coupling.\footnote{While the organic glass scintillator is chemically inert to silicone grease, it is partially soluble in it in a similar way that stilbene crystals and plastic scintillators are dissolved by some organic solvents. Due to the widespread use of silicone grease for optical coupling, methods to render the organic glass scintillator fully compatible with silicone grease have been developed concurrent to this work.} Both the silicone grease and the polyvinyl alcohol solution have near 100\% optical transmission over the emission spectra for the three organic scintillator materials~\cite{PVATransmission,EJ550spec}. To reflect the scintillation light, the EJ-276 and organic glass samples had all remaining sides wrapped in over ten layers of polytetrafluoroethylene tape~\cite{Janecek2012_Reflectivity}. The PMT response function was measured at each operating bias voltage using a finite difference method adapted from Friend et al.~\cite{friendLinearity} and pulse linearity was confirmed for each PMT-bias-voltage combination unless otherwise noted. 

The EJ-309 liquid cell contains an inert gas bubble, which comprises $\sim$3$\%$ of the chamber. If the bubble is located between the scintillating liquid and the photocathode of the PMT, a decrease in the light collection efficiency of $\sim$8\% has been observed. The results presented here are for configurations where the bubble was not located between the scintillating liquid and the photocathode.

For each study, full waveforms were acquired using a CAEN V1730 500~MS/s digitizer. Data reduction was accomplished using an object-oriented C\texttt{++} library incorporating elements of the ROOT data analysis framework~\cite{brun}. For the relative electron light output and the proton light yield studies, a mean waveform was obtained by averaging millions of pulses with energies spanning 25-75\% of the digitizer dynamic range. Integration lengths were set to achieve collection of over 95\% of the scintillation light in the average waveform emitted within a 600~ns acquisition window. Waveforms were reduced to pulse integrals using the integration lengths reported in table~\ref{LO-Res}. For the PSD studies, the integration length was treated as a free parameter. This is further discussed in section~\ref{PSD_Measurement}.


\section{Relative electron light output}
\label{ElectronLO}

A comparison of the relative electron light output of EJ-309, EJ-276, and the organic glass scintillator was realized by comparing the location of the Compton edge of a 662~keV $\gamma$ ray from a 4.2~$\mu$Ci~$^{137}$Cs source. Each of the scintillating media were coupled to a photodetector on one base using the same Hamamatsu Photonics H1949-51 PMT biased at $-1700$~V. The $^{137}$Cs source was positioned approximately 10~cm from the face of each scintillator to ensure uniform illumination. The channel number corresponding to the Compton edge of the 662~keV $\gamma$ ray was obtained by $\chi^2$ minimization between the measured spectrum and a Geant4 simulation \cite{Geant4} convolved with a resolution function~\cite{DIETZE}. The resolution of the pulse integral, $L$, is given by:
\begin{equation}
\label{resolutionEq}
\frac{\delta L}{L} = \sqrt{E_c^2+\frac{E_1^2}{L}+\frac{E_2^2}{L^2}}, 
\end{equation}
where $E_c$, $E_1$ and $E_2$ are fitting parameters. 

\begin{table*} 
	\caption{Relative electron light output (normalized to that of EJ-309), energy resolution, integration length, and measurement date for the three scintillating media. \label{LO-Res}}
	\centering
	\begin{tabular}{lcccc}
		\hline
		Scintillator  & Light Output & Energy Resolution & Integration Length & Measurement date\\ 
		&[ch] at 477 keV & [\%] at 477 keV & [ns] & \\ \hline   
		EJ-309 & 1.00  &  $10.7\pm0.4$ & 300 & September 2018 \\
		 & 1.00   & $11.0\pm0.2$ & 300 & March 2019 \\
		 & 1.00   & $11.4\pm0.4$ & 300 & October 2019 \\ \hline	
		EJ-276 & $0.79\pm0.02$   & $22.9\pm0.9$ & 350 & September 2018 \\
		 & $0.72\pm0.02$   & $26.0\pm0.6$ & 350 & March 2019 \\
		 & $0.62\pm0.02$   & $30.4\pm0.7$ & 350 & October 2019\\ \hline
		Organic Glass & $1.46\pm0.02$ & $13.4\pm0.4$ & 200 & March 2019 \\
		              & $1.45\pm0.02$ & $12.6\pm0.5$ & 200 & October 2019 \\
		\hline
	\end{tabular}
\end{table*}

\begin{figure}
	\center
	\includegraphics[width=0.90\textwidth]{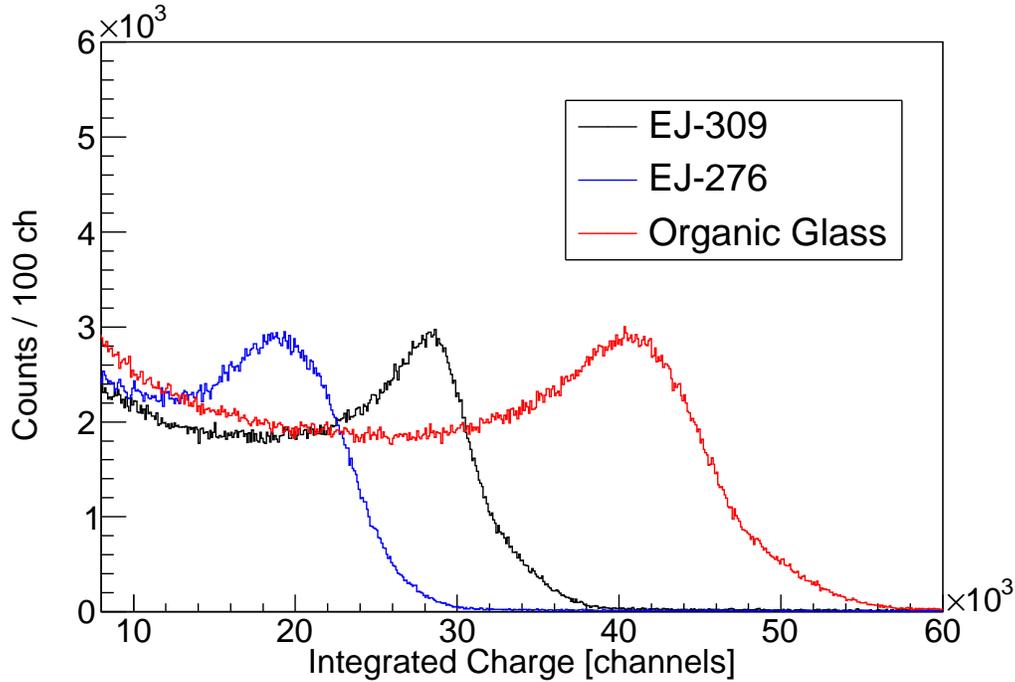}
	\caption{(Color Online) $^{137}$Cs pulse integral spectra for EJ-276 (blue), EJ-309 (black), and the organic glass (red), measured in March 2019. \label{relativeLightOutput}}
\end{figure}

Figure~\ref{relativeLightOutput} shows the pulse integral spectra obtained for each scintillator material. The position of the Compton edge relative to EJ-309, the resolution at the Compton edge energy (estimated using Eq.~\ref{resolutionEq}), the pulse integration length, and the date of the measurement for each material are reported in table~\ref{LO-Res}. The fit of the data to the Geant4 simulation was performed over the range of $0.3-0.6$~MeVee. The reported uncertainty is the standard deviation obtained by varying the limits of the fit range within 10\% of their nominal values over 200 trials. 

As shown in table~\ref{LO-Res}, the light output of the organic glass greatly exceeds that of EJ-309 and EJ-276. The resolution of the organic glass is slightly worse than that of EJ-309, despite the higher light output, though the EJ-276 scintillator demonstrated significantly poorer resolution. Rapid degradation of the light output of EJ-276 was also observed as a function of time. Zaitseva et al.\ previously measured the electron light output of a fresh EJ-276 sample as comparable to that of EJ-309~\cite{zaitseva2018}. The authors confirmed that after approximately six months the light output of the EJ-276 sample had decreased by $\sim30\%$~\cite{PrivCommZaitseva}. 

The potential for bias introduced by the wavelength-dependent photocathode quantum efficiency (QE) and the disparate scintillator emission spectra was investigated. The emission spectrum for EJ-309 was obtained from Eljen Technology~\cite{309spec}. Photoluminescence emission spectra of the EJ-276 and organic glass scintillator were measured using a setup featuring a broad-spectrum XS-433 Xe lamp as excitation source coupled to a SP-2150i monochromator (both from Acton Research Corporation). The emitted light was collected by a SpectraPro-2150i spectrometer and a PIXIS: 100B thermoelectrically-cooled charge-coupled device (CCD) (both from Princeton Instruments) with response in the $220-900$~nm range. Emission spectra were corrected to take into account the system response using the method of Derenzo et al.~\cite{Derenzo2008_IEEE, Derenzo2018_JAP}. 

\begin{figure}
	\center
	\includegraphics[width=0.90\textwidth]{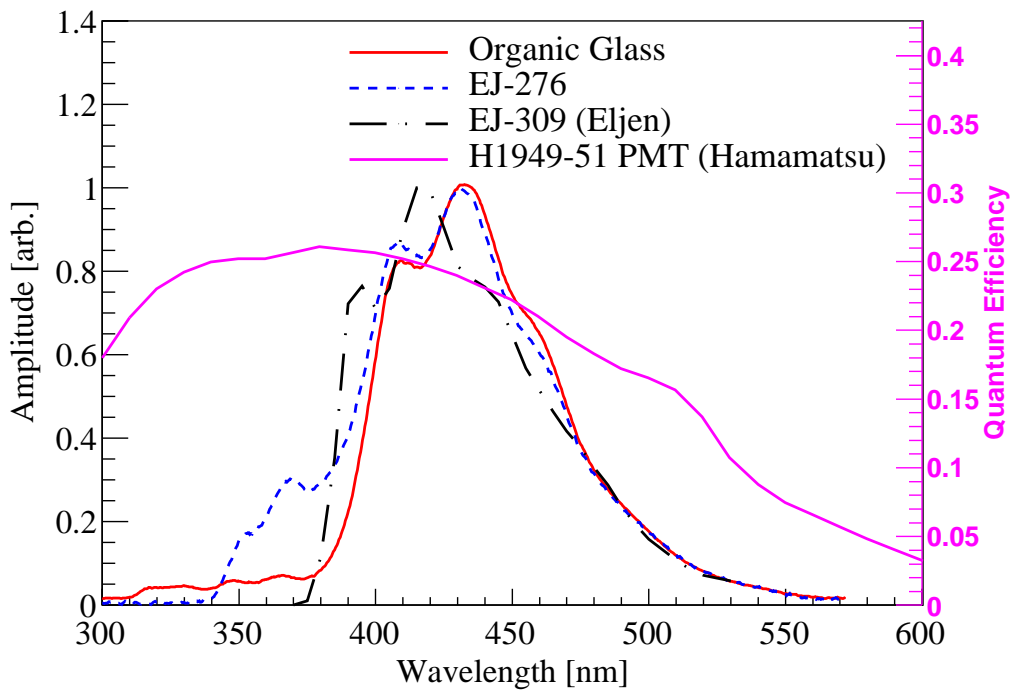}
	\caption{(Color Online) Emission spectra for the organic glass (red), EJ-276 (dashed blue), and EJ-309 (dashed black), the latter obtained from Eljen Technology. The emission spectra of the organic glass and EJ-276 were obtained by averaging emission spectra acquired with excitation wavelengths ranging from $240-290$~nm in 1~nm increments. The quantum efficiency of the H1949-51 PMT as reported by Hamamatsu Photonics is shown in pink, using the secondary ordinate axis.} \label{emissionSpectra}
\end{figure}

Figure~\ref{emissionSpectra} shows the emission spectra of EJ-276 and the organic glass obtained in this work, along with the emission spectrum of EJ-309 provided by Eljen Technology. The reported emission spectra correspond to the average of emission spectra measured for excitation wavelengths ranging from $240-290$~nm in 1 nm increments. The excitation source and the CCD were positioned facing the same side of the sample. As such, the emission response below $\sim 400$~nm corresponds in part to scintillation at the surface of the sample from the matrix (for the organic glass) and the primary dye (for EJ-276) \cite{Carlson2017_OGlassJACS, zaitseva2018}. Above $\sim400$~nm, the emission is similar for EJ-276 and the organic glass, whereas the emission spectrum of EJ-309 is slightly shifted to lower wavelengths.

The QE for the H1949-51 PMT assembly was provided by Hamamatsu Photonics and is reported in figure~\ref{emissionSpectra} on the secondary ordinate axis. To quantify the potential bias, 95 different excitation wavelengths were selected using the monochromator ranging from $230-400$~nm. The integrated product of the emission spectrum and the quantum efficiency curve was calculated for both EJ-276 and the organic glass for each excitation wavelength. The maximum change in the relative electron light output induced by the mismatch of the QE and emission spectra occurred at an excitation wavelength of 332~nm and corresponded to a less than 3\% effect with a standard deviation less than 1\%. Using the EJ-309 emission spectrum provided by Eljen Technology, the maximum change in the relative electron light output from this effect is less than 4\% between the three materials. 


\section{Proton light yield}
\label{exp}
\subsection{Experimental methods}
Proton light yield measurements were carried out at the 88-Inch Cyclotron at Lawrence Berkeley National Laboratory using the method of Brown et al.~\cite{brown-thesis,brown2018proton}. Neutrons were produced by impinging a $^2$H$^+$ beam onto a metal target located in the cyclotron vault~\cite{harrig}. The target scintillator to be characterized was placed in-beam, an array of PSD-capable liquid scintillators was placed out-of-beam, and coincident events in the target and observation detectors were obtained. The incoming neutron energy, E$_{n}$, and outgoing neutron energy, E$_{n'}$, were determined using the measured time-of-flight between the breakup target and the target detector and between the target detector and one of the observation detectors, respectively. For n-p elastic scattering events in the target, the energy of the recoiling proton, E$_{p}$, was calculated using the incoming neutron energy and the scattering angle, $\theta$, with respect to the incoming neutron direction:
	\begin{linenomath*}
		\begin{equation}
		\label{thisone}
		E_p = E_n \sin^2\!{\theta}. 
		\end{equation}
	\end{linenomath*} 
The proton recoil energy associated with a given target scintillator pulse was determined on an event-by-event basis. Measurement of the scattered neutron energy was used to match the incoming neutron with the corresponding cyclotron beam pulse~\cite{brown2018proton}.

\begin{figure}
	\centering
	\includegraphics[width=0.9\textwidth]{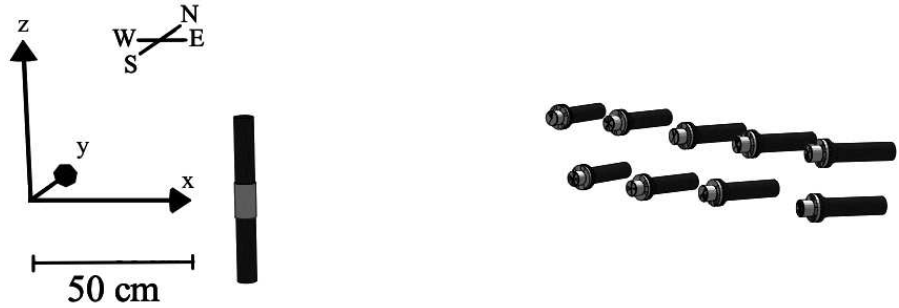}
	\caption{Schematic of the detector array configuration for the EJ-276 proton light yield measurement. The neutron beam traveled along the $x$-axis through the target scintillator with dual-ended readout with observation detectors positioned at forward angles with respect to the incoming beam. Similar configurations were used for the EJ-309 and organic glass proton light yield measurements.}
	\label{EJ276ExpSetup}
\end{figure}

A schematic of the experimental setup for the EJ-276 proton light yield measurement is shown in figure~\ref{EJ276ExpSetup}; the EJ-309 and organic glass measurements used similar configurations. Appendix~\ref{appDetLoc} gives the detector locations for the EJ-309, EJ-276, and organic glass proton light yield measurements, the latter of which involved two different configurations for low and high target bias voltage settings covering proton recoil energies from 800~keV to 20~MeV and 50~keV to 2.3~MeV, respectively. The target scintillator volumes and treatment are described in section~\ref{materials}. The EJ-309 target scintillator was coupled on one base to a Hamamatsu Photonics H1949-50 PMT, whereas the EJ-276 and organic glass target scintillators were coupled on both bases to Hamamatsu H1949-51 PMTs in a similar dual-window configuration as described in~\cite{EJ20XPLY}. For each experiment, the observation detectors (shown at forward angles with respect to the target scintillator in figure~\ref{EJ276ExpSetup}) were composed of 5.08 cm dia.\ $\times$ 5.08 cm length right circular cylindrical EJ-309 cells contained in a thin aluminum housing and mounted to either a Hamamatsu H1949-50 or H1949-51 PMT on one end. The optical coupling between the observation scintillators and the borosilicate glass window of the PMTs was achieved using BC-630 silicone grease. The PMTs used for the target and observation scintillators were powered using CAEN NTD1470 and CAEN R1470ETD high voltage power supplies, respectively. 

The EJ-309 measurement employed a 16~MeV $^2$H$^+$ beam on a $3.8$-mm-thick Ta  target. Data were acquired over a period of approximately 8~h with a beam current of $\sim1.6$~$\mu$A. The EJ-276 measurement also used a 16~MeV $^2$H$^+$ beam, but on a $3$-mm-thick Be target. Data were acquired over a period of approximately 9~h with a beam current of $\sim90$~nA. The lower beam current for the EJ-276 measurement compared to that used for EJ-309 reflects the different neutron production cross sections for deuterons incident on Ta and Be~\cite{meulders1975fast}. The organic glass measurement used a 33~MeV $^2$H$^+$ beam on a $3$-mm-thick Be breakup target. Two measurements with different PMT bias voltages were realized to cover a broad range of proton recoil energies. Data were acquired over a period of approximately 8~h with a beam current of $\sim 35$~nA for the low gain setting and approximately 11 hours with a beam current of $\sim10$~nA for the high gain setting. 

For the EJ-309 measurement, the master trigger required a coincidence between at least one observation detector and the target scintillator detector. For the EJ-276 and organic glass measurements, the trigger required a coincident signal between both PMTs coupled to the target scintillator and at least one observation detector. The RF control signal from the cyclotron, providing the beam pulse frequency, was recorded with each trigger. Trigger time tags were established using the CAEN digital constant fraction discrimination (CFD) algorithm with a 75\% fraction and 4~ns delay for the scintillator signals. The RF trigger time tags were established using the same CFD algorithm with a 75$\%$ fraction and 24~ns delay for the EJ-309 measurement, and leading edge discrimination for the EJ-276 and organic glass measurements~\cite{DPP-PSD}. 

\subsection{Signal processing and light output calibration}
\label{LOCalibration}

Table~\ref{expDetailsTable} provides the target detector bias voltage settings for each measurement. The lower bias voltages used for the organic glass measurement exhibited a nonlinear PMT response and raw traces were corrected on a sample-by-sample basis as described in ref.~\cite{brown-thesis}. The impact of the PMT nonlinearity correction was less than 4\% over the full energy range of the measurement. 

\begin{table} 
	\caption{Summary of the target PMT bias voltage settings used for the different scintillating media. \label{expDetailsTable}}
	\centering
	\begin{tabular}{ccccc}
		\hline
	        & EJ-309 & EJ-276 & Glass & Glass \\ 
	        & & & (low gains) & (high gains) \\
	        \hline
		Upper PMT bias [V]&$-1750$&$-1675$&$-1350$&$-1913$\\ 
		Lower PMT bias [V]&N/A&$-1760$&$-1282$&$-1800$\\ 
		\hline
	\end{tabular}
\end{table}

For the EJ-276 and the organic glass scintillators, the two target PMTs were gain-matched using a $^{137}$Cs source located approximately 10~cm from the center of the scintillator to ensure uniform distribution of electron recoils along the length of the cylinder. The pulse integral values obtained from each target PMT were combined using the geometric average to obtain the total scintillation light independent of the interaction location~\cite{Knollch10}. 

The traditional MeVee electron-equivalent light unit assumes a linear response of the electron light yield and is not a useful quantity for scintillators that exhibit significant nonlinearity. Rather than relying upon the (potentially unjustified \cite{payne}) assumption of a linear electron response over the full range of the measurements, the light unit in this work was determined relative to that of a 477~keV electron using the Compton edge of a 662 keV $\gamma$ ray from a $^{137}$Cs source, i.e., a light unit of 1 corresponds to the amount of light produced by a 477 keV recoil electron. The electron light output calibration for all three materials proceeded as described in section~\ref{ElectronLO}.

\subsection{Timing calibrations}
\label{TOFCalib}

\begin{figure}
	\center
	\includegraphics[width=0.9\textwidth]{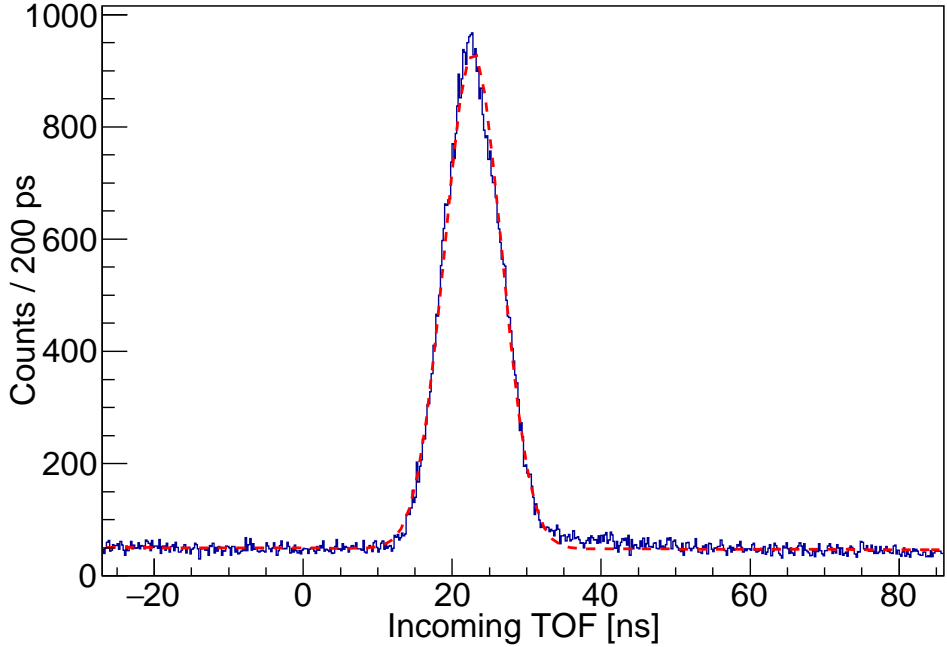}
    	\caption{(Color online) The blue curve shows a histogram of time differences between the cyclotron RF signal and $\gamma$-ray events in the EJ-309 target detector. The width of the pulse is primarily reflective of the spatial spreading of the beam pulse, which is the dominant source of systematic uncertainty in the incoming TOF determination, with $\sigma=3.76$~ns. The red dashed curve is a fit of the measured data with a normal distribution plus linear background term.} \label{incTOF}
\end{figure}

Timing calibrations were performed for both the incident and outgoing time-of-flight (TOF) measurements as described in ref.~\cite{brown2018proton}. In cases where the target scintillator was coupled to two PMTs, the time of the target event was taken as the average of the clock times for the two PMTs as described in ref.~\cite{EJ20XPLY}. For the incoming TOF calibration, a histogram of time differences between the cyclotron RF signal and $\gamma$-ray events in the target detector was constructed, as illustrated for the EJ-309 target scintillator in figure~\ref{incTOF}. Similarly, for the outgoing TOF calibration, histograms of time differences between $\gamma$-ray interactions in the target and observation detectors were obtained. A binned maximum likelihood estimation between the photon interaction feature and the superposition of a normal distribution with a linear background was performed to obtain the centroid of the distribution and the associated statistical error, which were then used to calibrate the signal time chain. The standard deviations of the fitted normal distributions for the incoming and outgoing TOF calibrations are given in appendix~\ref{appDetLoc}.

\subsection{Proton energy uncertainty}
\label{dataReduction}

\begin{figure}
	\center
	\includegraphics[width=0.9\textwidth]{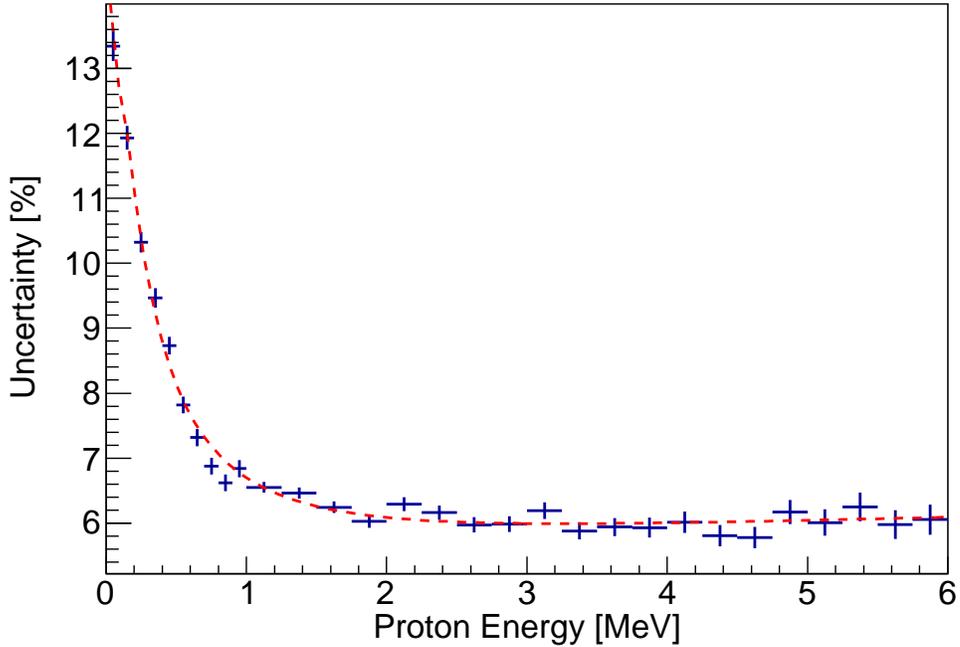}
	\caption{(Color online) Estimation of the proton energy resolution obtained via parametric bootstrapping. The red dashed line is a rational fit to the results used to set the proton energy bin width. \label{protonEnergyResolutionWFit}}
\end{figure}

The uncertainty in the measured proton recoil energy is dependent upon the uncertainty in the incoming TOF (dominated by the spatial spreading of the cyclotron beam pulse) and the angular and pathlength variance arising from the interaction locations in both the target and observation detectors. A synthetic data set was generated via Monte Carlo sampling using Geant4 to account for the variance in proton recoil energies due to the finite size of the scintillators \cite{Geant4}. Using parametric bootstrapping, each simulated event was processed adding uncertainty sampled from a normal distribution of the experimentally-determined incoming TOF uncertainty. The proton recoil energy was then calculated assuming that the interaction had taken place at the center of the cells, as in the experimental data analysis. The calculated proton recoil energy was compared to the proton recoil energy observed in the Geant4 simulation. The distributions of the difference between the calculated and ``true'' proton recoil energies as a function of proton recoil energy were fitted with normal distributions. The standard deviation as a function of proton recoil energy is shown in figure~\ref{protonEnergyResolutionWFit}, along with a rational function fit to the data used to set the proton energy bins. 

\begin{figure*}
	\centering
	\begin{subfigure}{0.32\textwidth}
		\centering
		\includegraphics[width=0.97\textwidth]{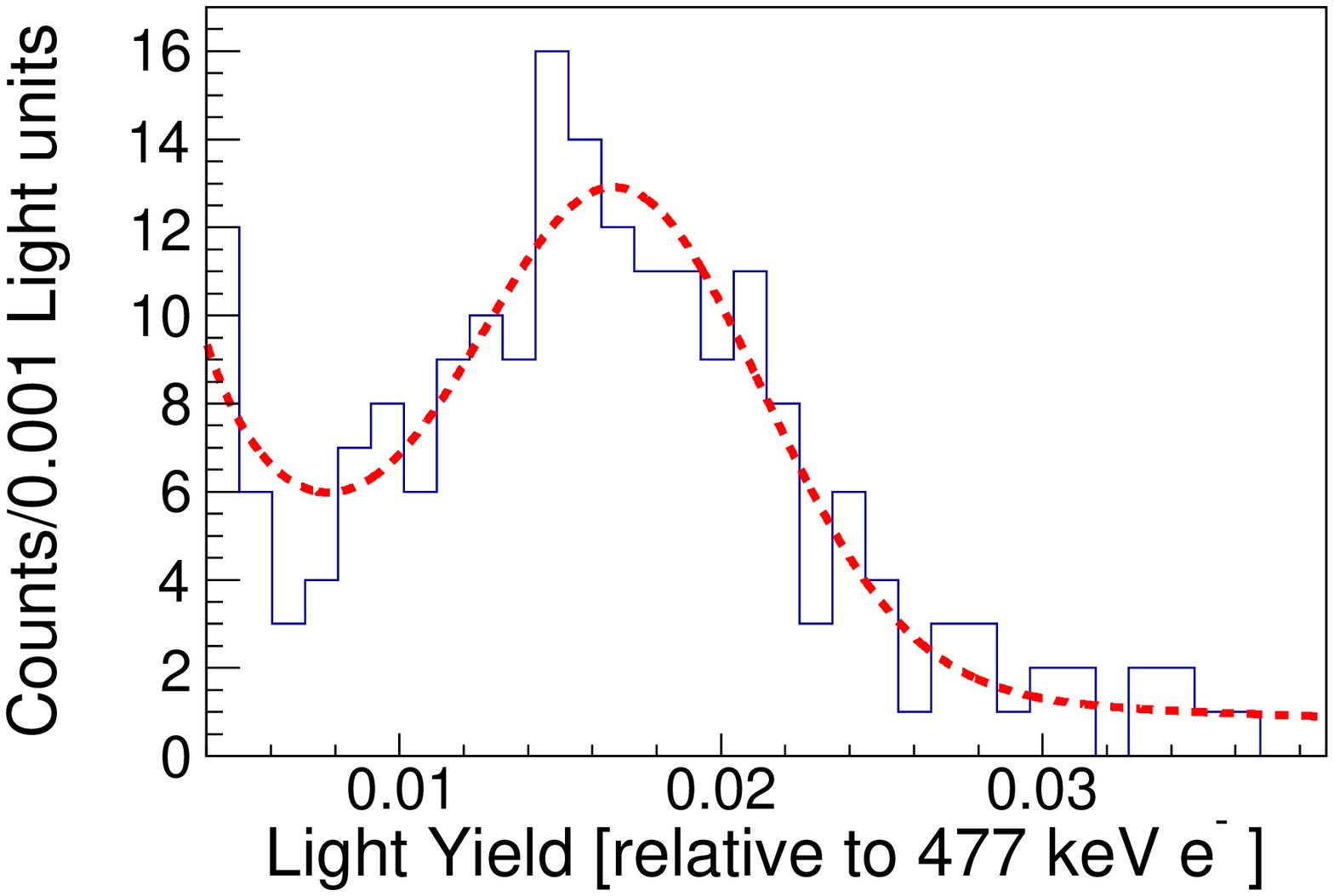}
		\caption{$\overline{\mathrm{E}_\mathrm{p}}$ = 64 keV. \label{lowest}}
	\end{subfigure}%
	~ 
	\begin{subfigure}{0.32\textwidth}
		\centering
		\includegraphics[width=0.97\textwidth]{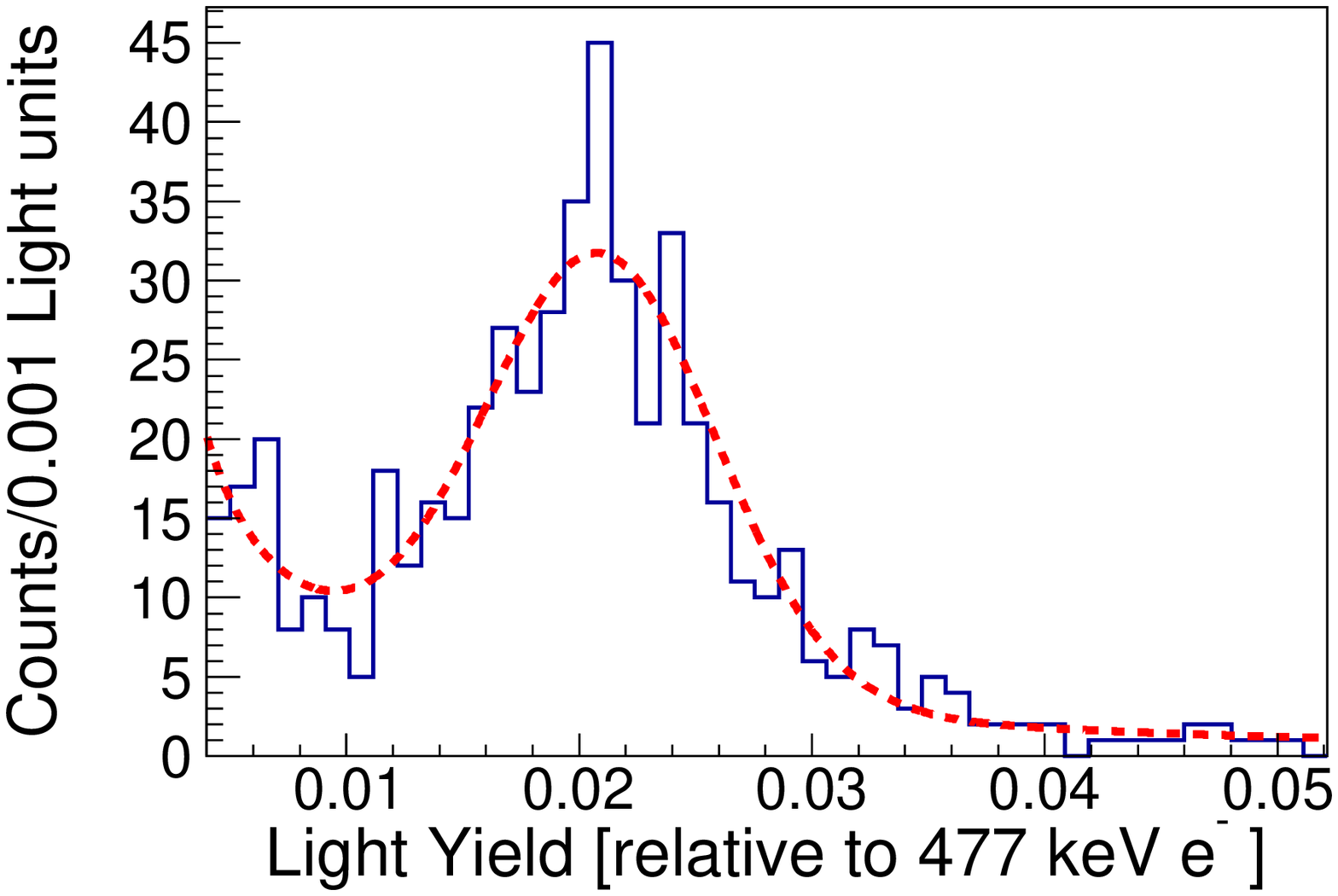}
		\caption{$\overline{\mathrm{E}_\mathrm{p}}$ = 88 keV.}
	\end{subfigure}
	~
	\begin{subfigure}{0.32\textwidth}
		\centering
		\includegraphics[width=0.97\textwidth]{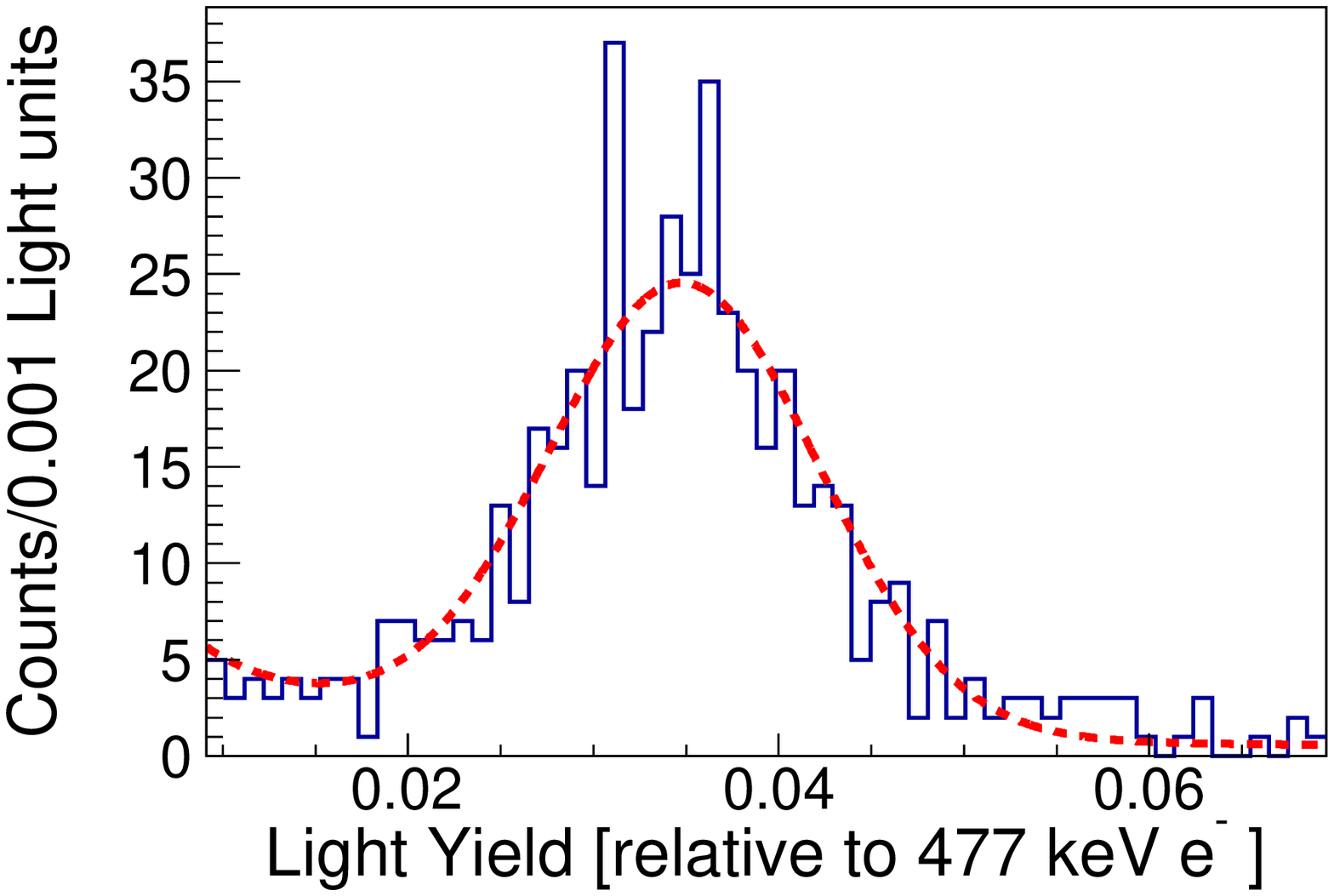}
		\caption{$\overline{\mathrm{E}_\mathrm{p}}$ = 193 keV.}
	\end{subfigure}
	
	\begin{subfigure}{0.32\textwidth}
		\centering
		\includegraphics[width=0.97\textwidth]{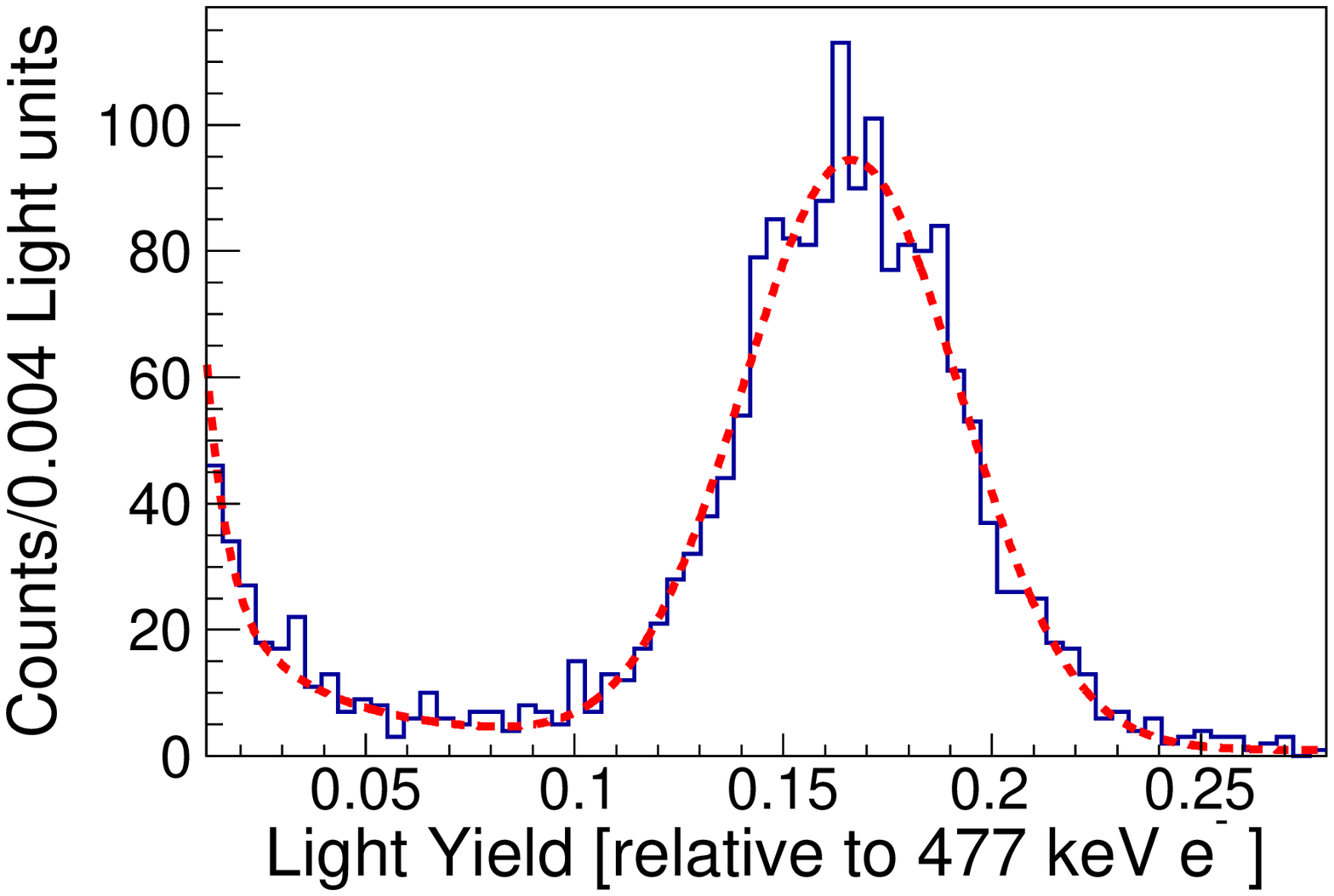}
		\caption{$\overline{\mathrm{E}_\mathrm{p}}$ = 478 keV.}
	\end{subfigure}%
	~ 
	\begin{subfigure}{0.32\textwidth}
		\centering
		\includegraphics[width=0.97\textwidth]{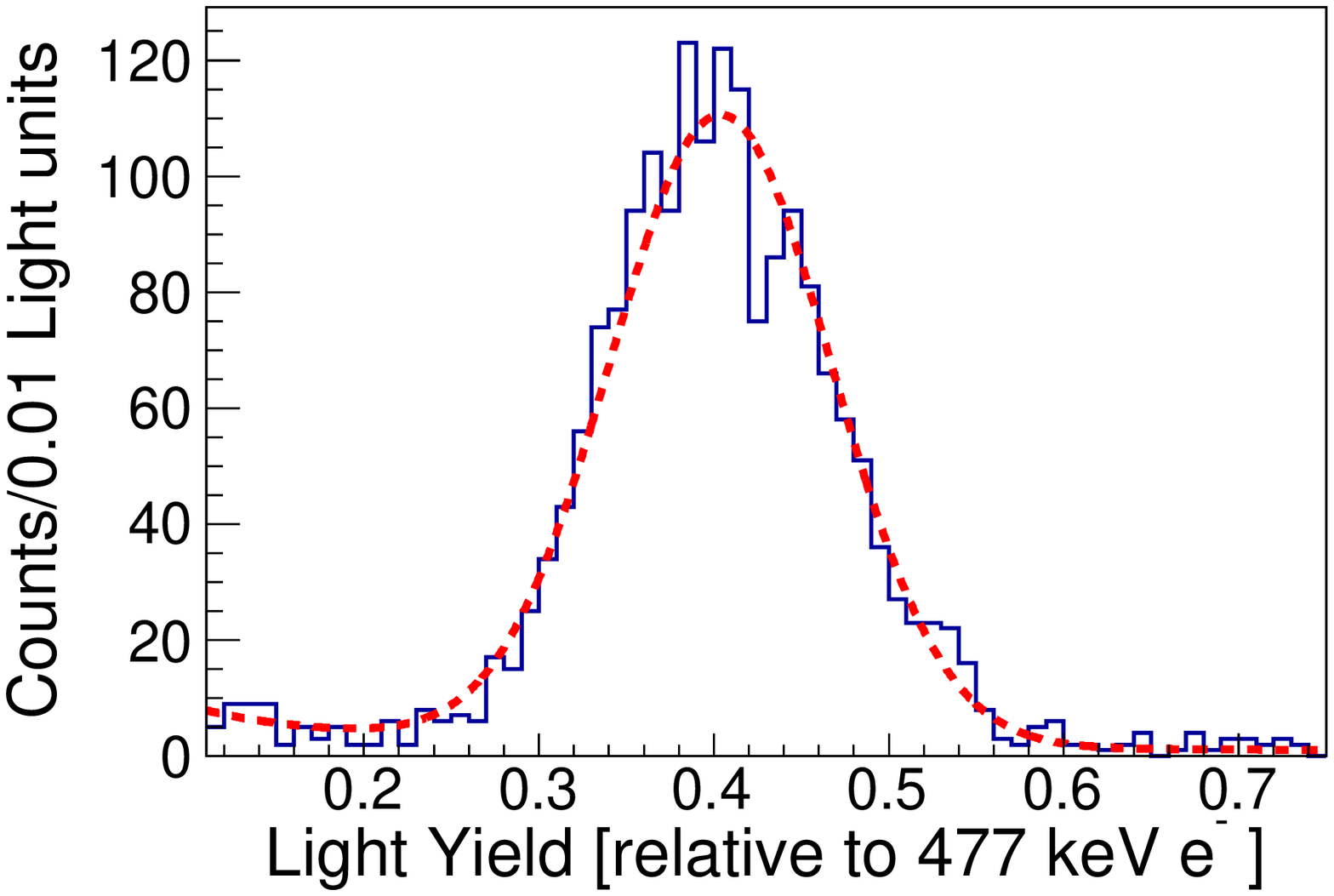}
		\caption{$\overline{\mathrm{E}_\mathrm{p}}$ = 870 keV.}
	\end{subfigure}
	~ 
	\begin{subfigure}{0.32\textwidth}
		\centering
		\includegraphics[width=0.97\textwidth]{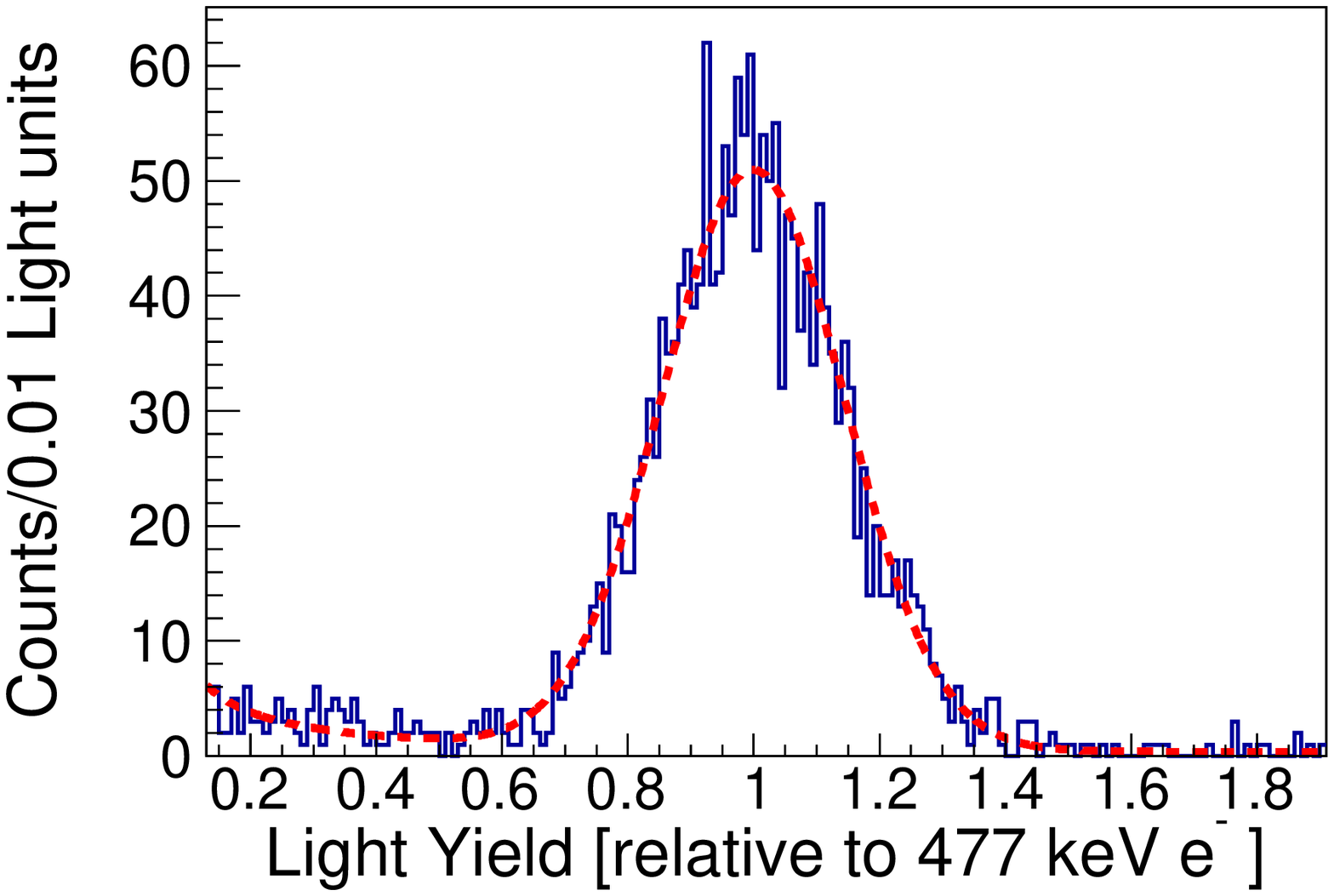}
		\caption{$\overline{\mathrm{E}_\mathrm{p}}$ = 1.59 MeV.}
	\end{subfigure}

    \begin{subfigure}{0.32\textwidth}
    	\centering
    	\includegraphics[width=0.97\textwidth]{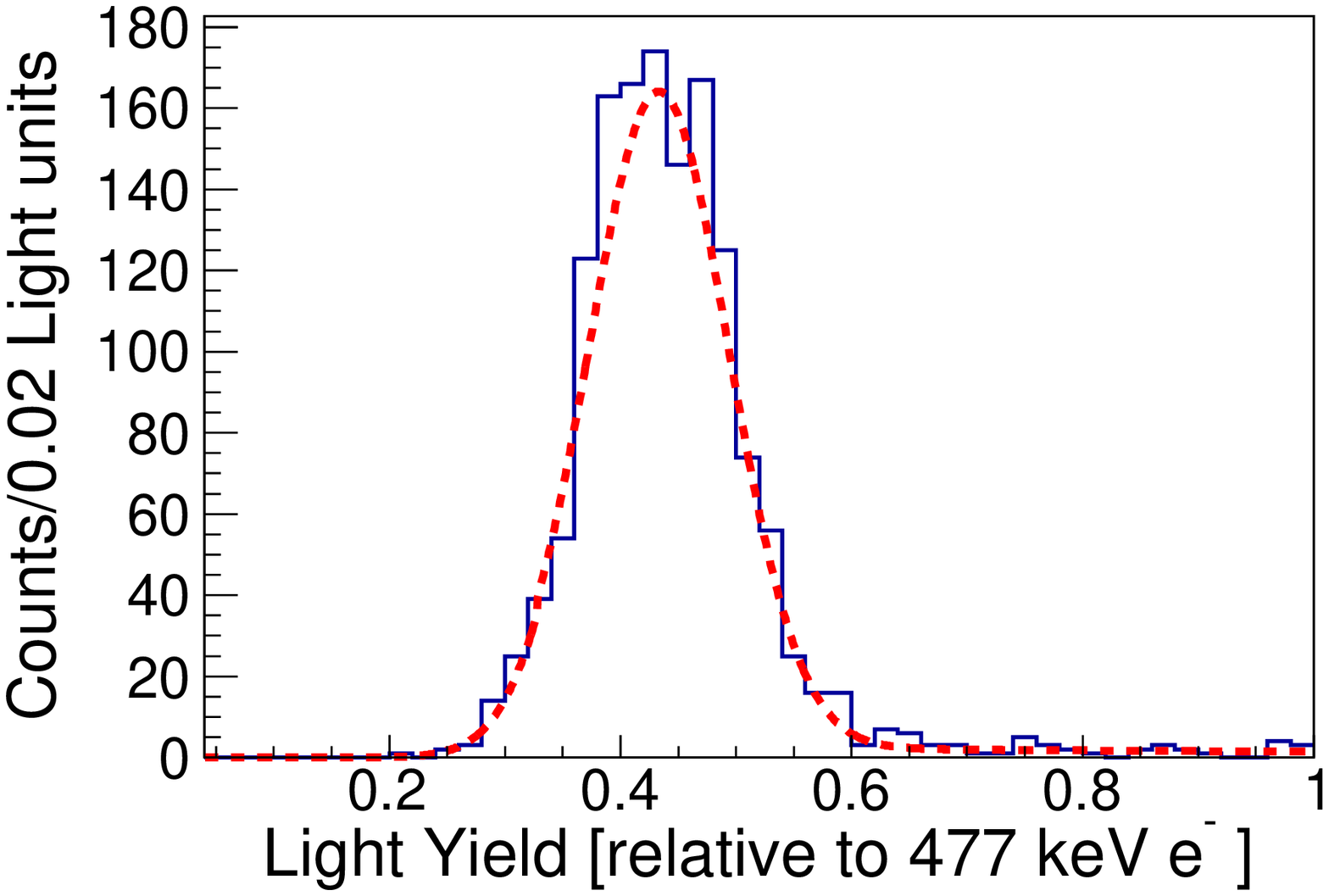}
    	\caption{$\overline{\mathrm{E}_\mathrm{p}}$ = 935 keV.}
    \end{subfigure}%
    ~ 
    \begin{subfigure}{0.32\textwidth}
    	\centering
    	\includegraphics[width=0.97\textwidth]{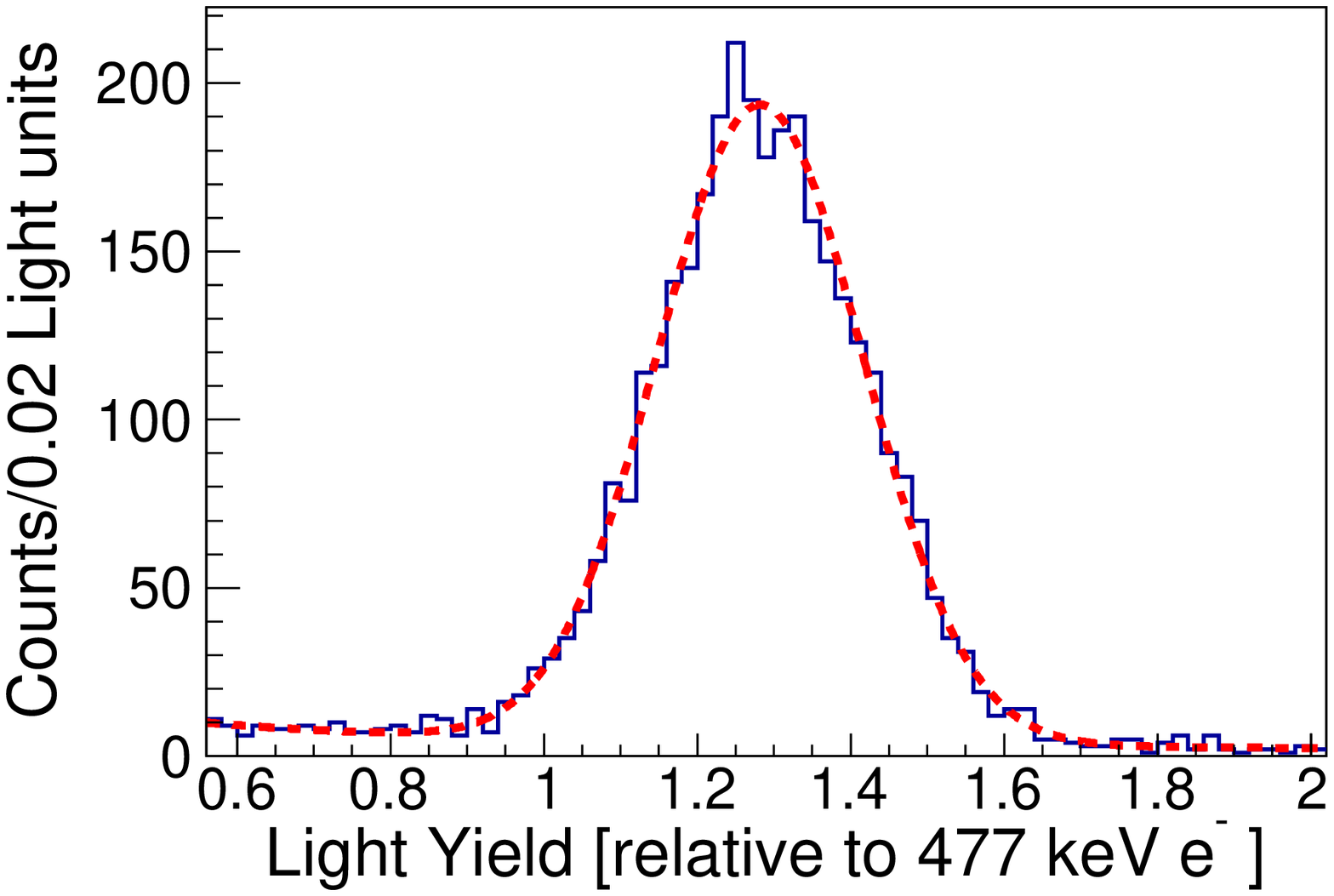}
    	\caption{$\overline{\mathrm{E}_\mathrm{p}}$ = 1.92 MeV.}
    \end{subfigure}
    ~ 
    \begin{subfigure}{0.32\textwidth}
    	\centering
    	\includegraphics[width=0.97\textwidth]{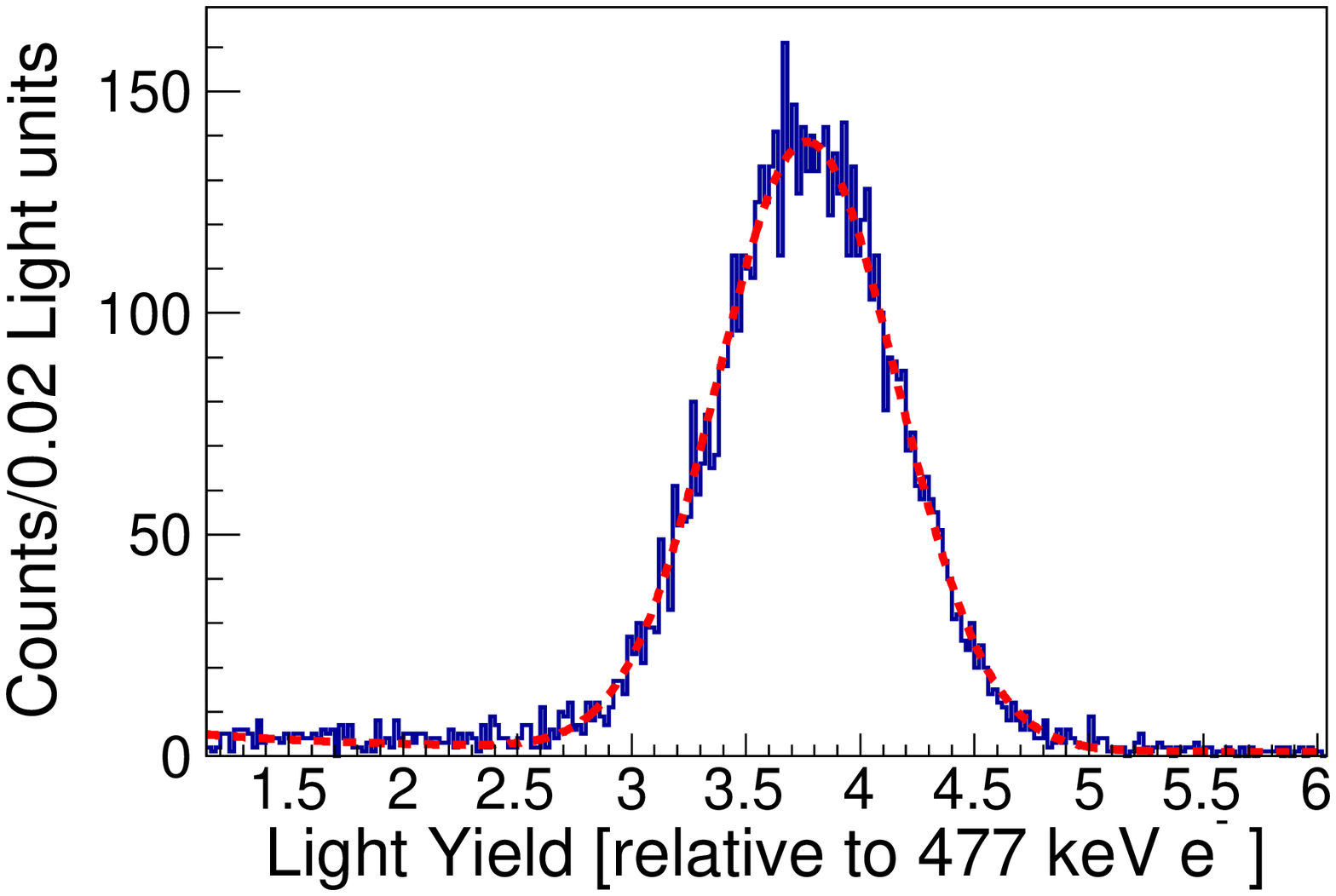}
    	\caption{$\overline{\mathrm{E}_\mathrm{p}}$ = 4.06 MeV.}
    \end{subfigure}
    ~ 
\begin{subfigure}{0.32\textwidth}
	\centering
	\includegraphics[width=0.97\textwidth]{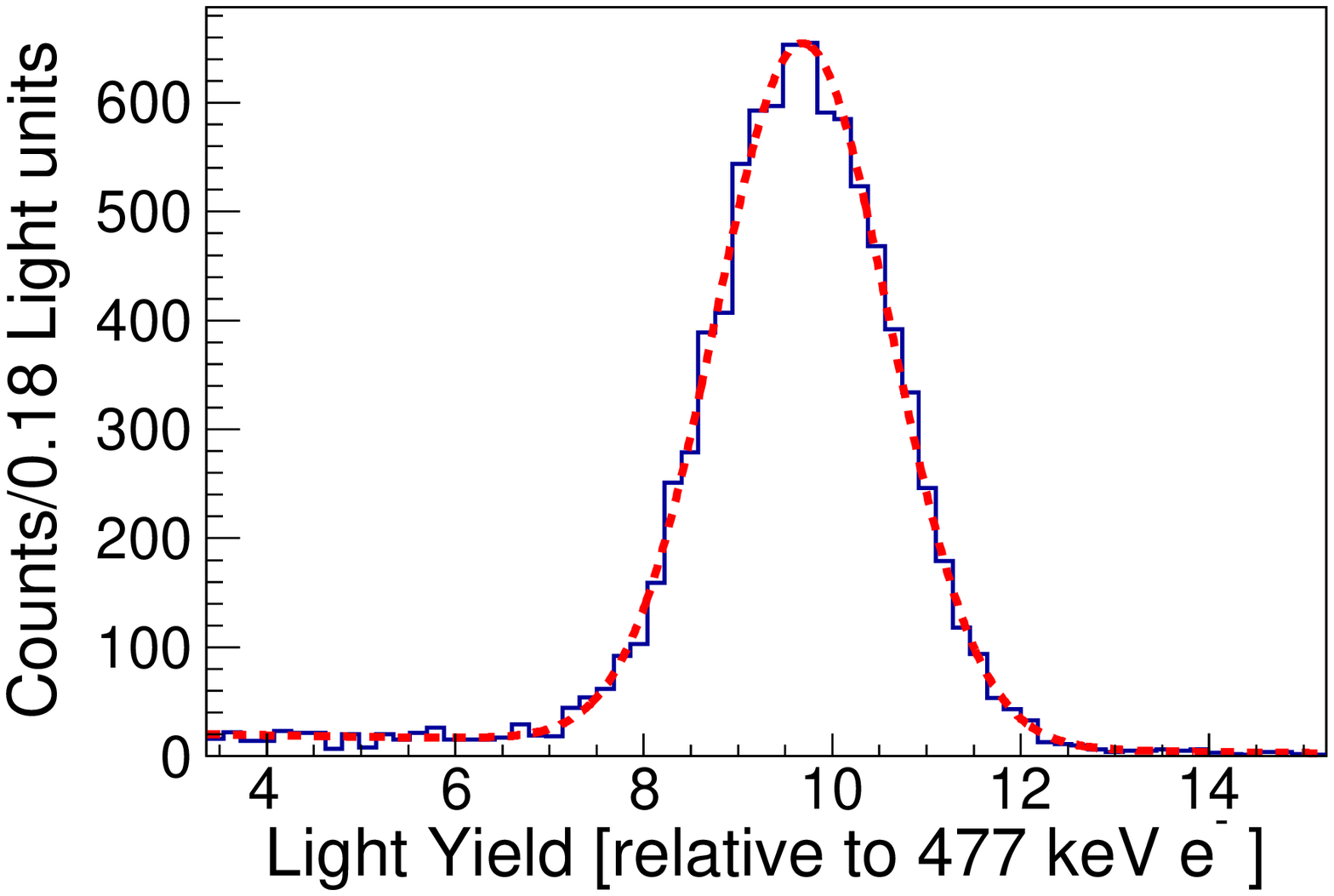}
	\caption{$\overline{\mathrm{E}_\mathrm{p}}$ = 8.10 MeV.}
\end{subfigure}
    ~ 
\begin{subfigure}{0.32\textwidth}
	\centering
	\includegraphics[width=0.97\textwidth]{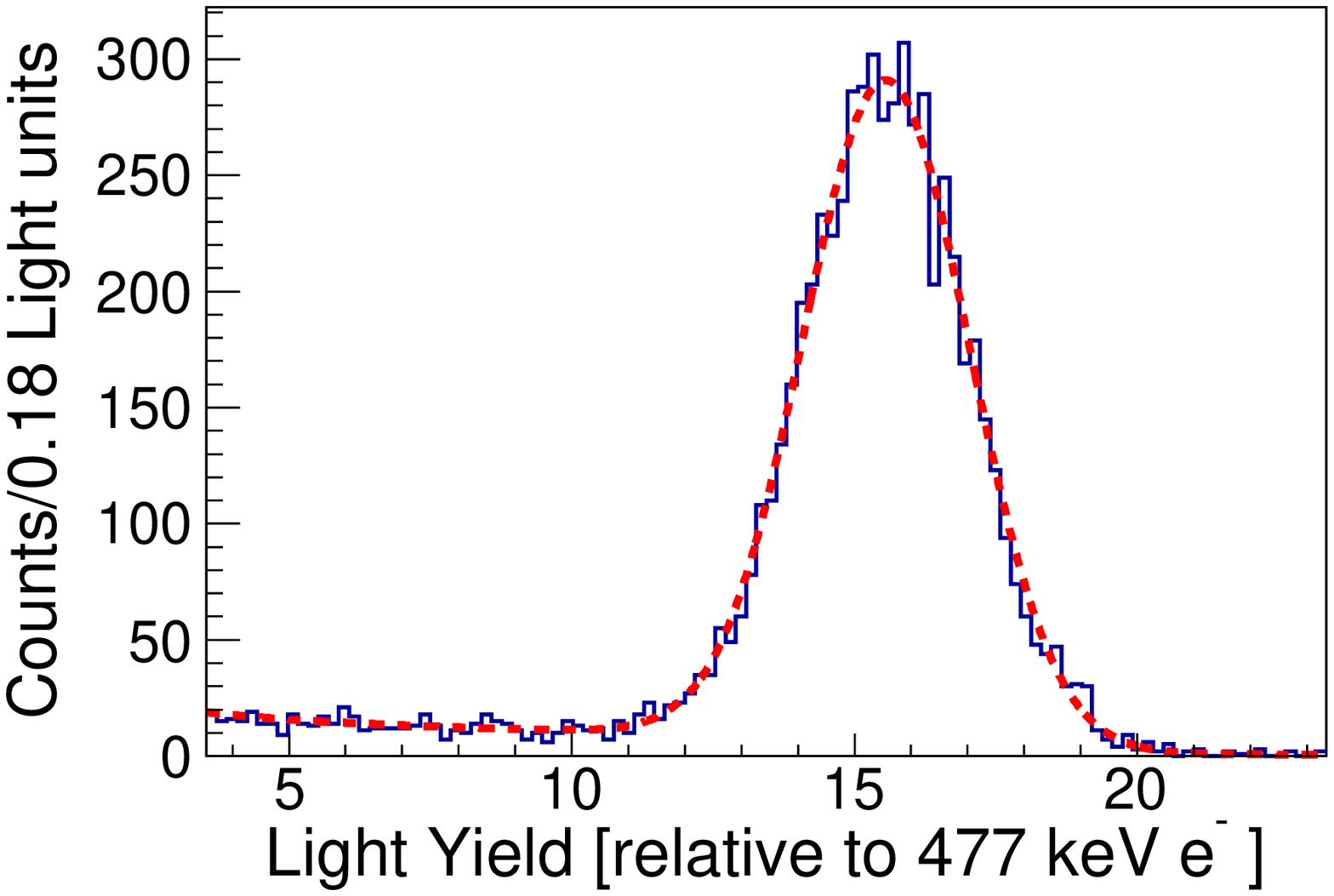}
	\caption{$\overline{\mathrm{E}_\mathrm{p}}$ = 11.7 MeV.}
\end{subfigure}
    ~ 
\begin{subfigure}{0.32\textwidth}
	\centering
	\includegraphics[width=0.97\textwidth]{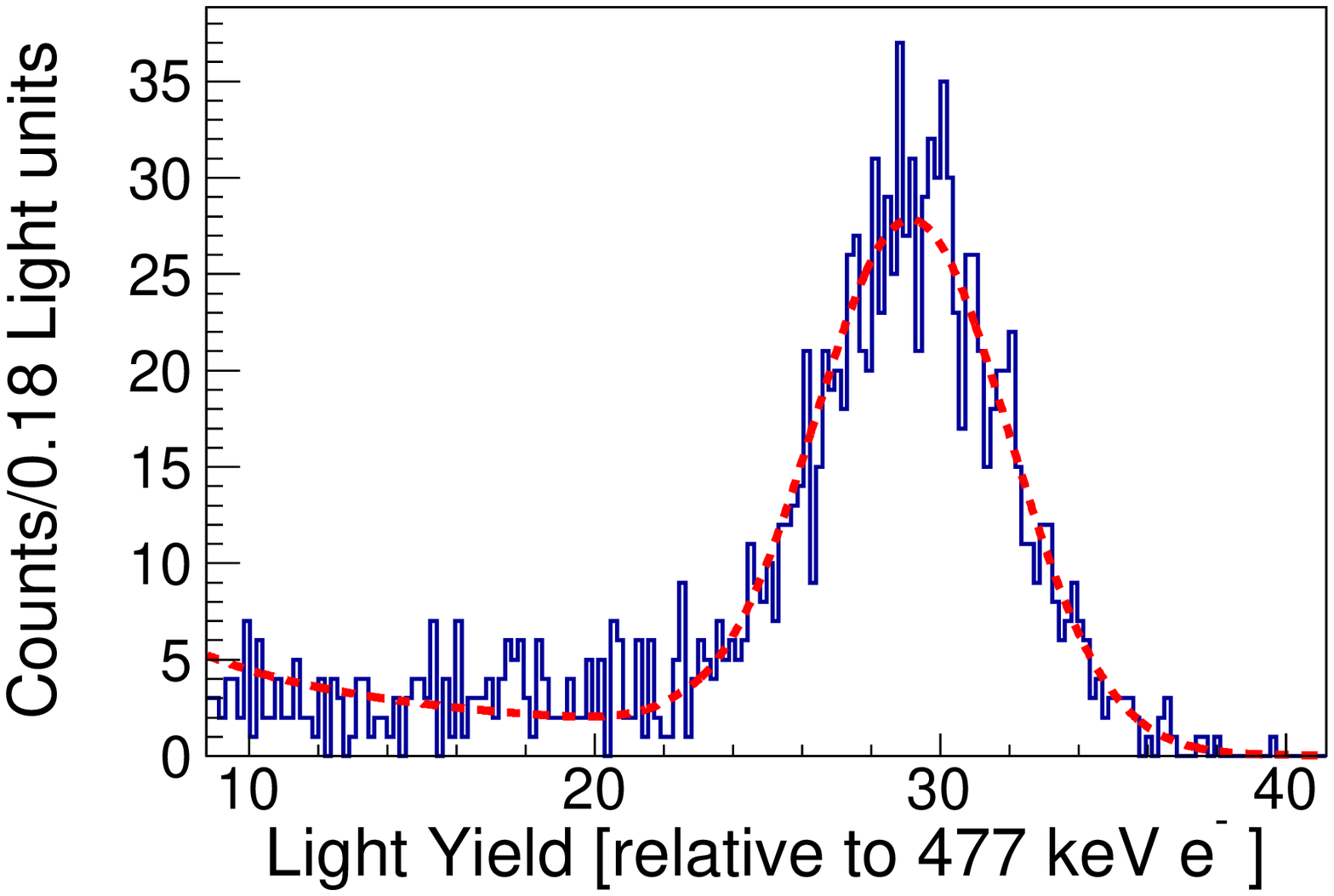}
	\caption{$\overline{\mathrm{E}_\mathrm{p}}$ = 19.4 MeV.}
\end{subfigure}
	\caption{(Color online) Organic glass relative light output spectra (blue curve) fit with a piecewise power law and a normal distribution (red dashed curves). Each spectrum corresponds to a different proton recoil energy bin with average energy, $\overline{\mathrm{E}_\mathrm{p}}$. The centroid of the normal distribution corresponds to the mean light production for n-p elastic scattering events within the bin. Spectra (a) through (f) correspond to data taken with the PMTs biased at high gains. Spectra (g) through (l) correspond to data taken with the PMTs biased at low gains to capture higher energy recoil protons.
	\label{sliceFit}}
\end{figure*}

\subsection{Light output determination and uncertainty}

Individual proton energy bins were projected onto the light output axis to produce a series of data points. The resulting histograms were fit using a binned maximum likelihood estimation considering the peak to be normally distributed and the background to be represented by a continuous cusped piecewise power law with two subdomains. Representative fits of the relative light output spectra for the organic glass are shown in figure~\ref{sliceFit}.
 
Several potential sources of systematic bias were identified, including the uncertainty in the detector positions, uncertainty in the flight path from the neutron production target to the origin of the coordinate system, and uncertainty in the incoming and outgoing TOF calibrations. To address these, a Monte Carlo approach was implemented where these inputs were varied assuming normally distributed uncertainties given in appendix~\ref{appDetLoc}. The reduction from histogrammed quantities to data points was repeated for each trial. The uncertainty in the channel number corresponding to the $^{137}$Cs edge determination, obtained by varying the fit range about the region of interest, was quantified at a $1\%$ maximum contribution. The PMT gain variation was evaluated by analyzing the dataset in batches corresponding to different time windows, and was evaluated to be $2\%$ for the EJ-309 and EJ-276 measurements and $1\%$ for the organic glass measurements. Those two contributions to the uncertainty were added in quadrature to the uncertainty obtained from the Monte Carlo simulation to provide the total uncertainty on the relative proton light yield. 

\subsection{Results and discussion}
\label{results}

\begin{figure}
	\center
	\includegraphics[width=0.9\textwidth]{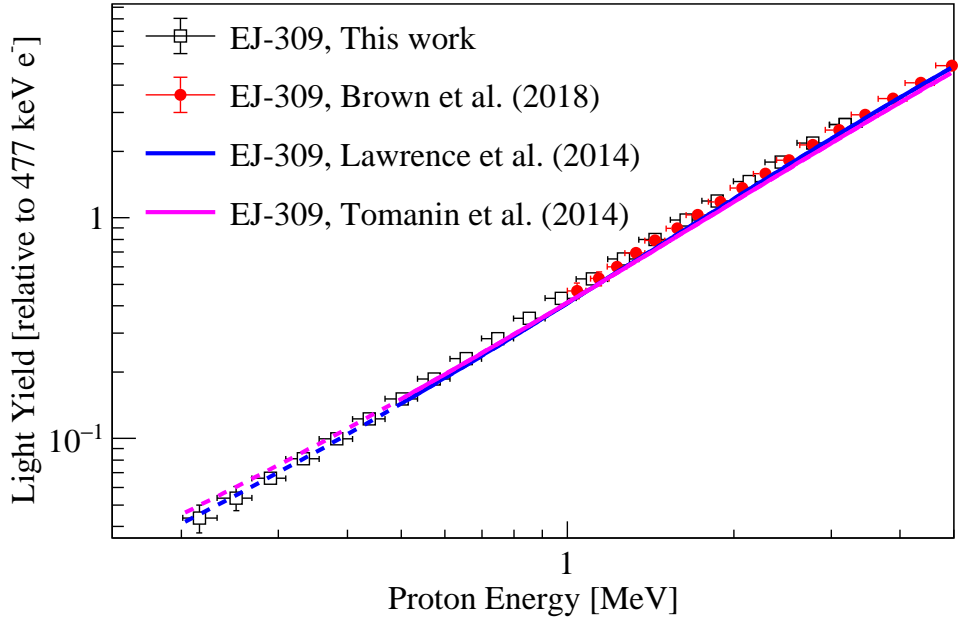}
	\caption{(Color online) The EJ-309 proton light yield, where the present work corresponds to the open squares. The same double time of flight method was used by Brown et al.\ at higher proton recoil energies (red circles). The $x$-error bars represent a bin width and the $y$-error bars include both the statistical and systematic uncertainty. The functional forms from Lawrence et al.\ (blue line) and Tomanin et al.\ (magenta line) are continuous over the measurement ranges and dashed in the extrapolation regions. \label{PLY309}}
\end{figure}

Figure~\ref{PLY309} shows the light yield of EJ-309 as function of proton recoil energy obtained in this work, along with several literature measurements. The present work (open squares) covers a proton recoil energy range of 200~keV to 3~MeV. For the proton light yield measurements presented here, the $x$-error bars represent a bin width, set using the proton energy resolution as described in section~\ref{dataReduction}, and the $y$-error bars include both the statistical and systematic uncertainties. The red circles were adapted from Brown et al.~\cite{brown2018proton}, having been re-calibrated to the current light unit using the raw experimental data. Both the present results and the work from Brown et al.\ used a double TOF technique with a 300~ns integration length and are in agreement within the estimated uncertainties over the common energy range. Lawrence et al.\ (blue curve) used an edge characterization technique and a 180~ns integration window~\cite{lawrence}. The proton light yield was measured in an energy range from 450~keV to 5.95~MeV and is approximately $4-8\%$ lower than this work. Tomanin et al.\ (magenta line) also used an edge characterization technique with a 120~ns integration window and measured the proton light yield from 500~keV to 14~MeV \cite{tomanin}. The results from Tomanin et al.\ are within $10\%$ of the present work. Based on average waveforms obtained in the proton recoil energy range of approximately 500~keV to 2~MeV and using a 700~ns acquisition window, the amount of light collected using an integration window of 180~ns and 120~ns is estimated to correspond to approximately 90\% and 86\% of the total scintillation light, respectively. For Lawrence et al.\ and Tomanin et al., the light unit conversion was performed presuming 0.477~MeVee corresponds to one light unit as both authors assumed electron light linearity. Both works provide empirical fit functions to characterize their measurements and these are used to represent their results in figure~\ref{PLY309}. The continuous line represents the energy range where the authors have measured data, and the dashed line represents an extrapolation of these functional forms over the range of this measurement.  

\begin{figure}
	\center
	\includegraphics[width=0.9\textwidth]{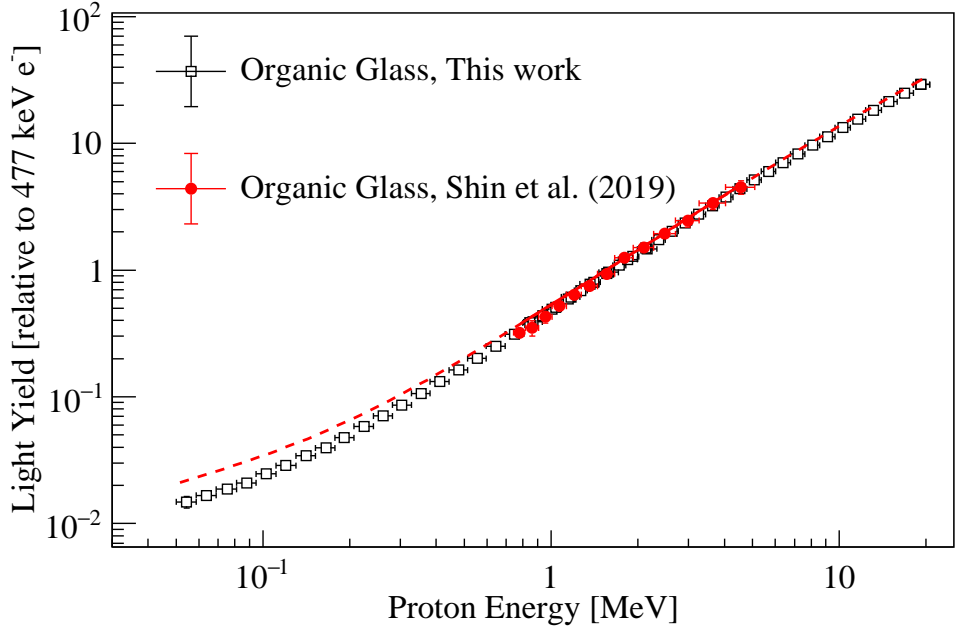}
	\caption{(Color online) Organic glass proton light yield. For this work, the $x$-error bars represent a bin width and the $y$-error bars include both the statistical and systematic uncertainty. The functional form from Shin et al.\ (red line) is continuous over their measurement range and dashed in the extrapolation regions. \label{PLYOG}}
\end{figure}

Figure~\ref{PLYOG} shows the organic glass proton light yield measured in this work (open squares) from 50~keV to 20~MeV. The data are within uncertainties of pulse integral measurements from Shin et al.~\cite{ShinOGlass2019} (red circles), which cover energies ranging from 800 keV to 4.6 MeV. The solid red line corresponds to the Birks functional form with fit parameters provided by Shin et al.\ over the measured energy range \cite{birks1964}. The extrapolation to lower and higher energies is shown by the dashed red line, which deviates at low energies from the measured data. Attempts to fit the full energy range of the data obtained in this work using the Birks relation yielded best fit results with part of the range outside $2\sigma$ of the estimated uncertainties.

\begin{figure}
	\center
	\includegraphics[width=0.9\textwidth]{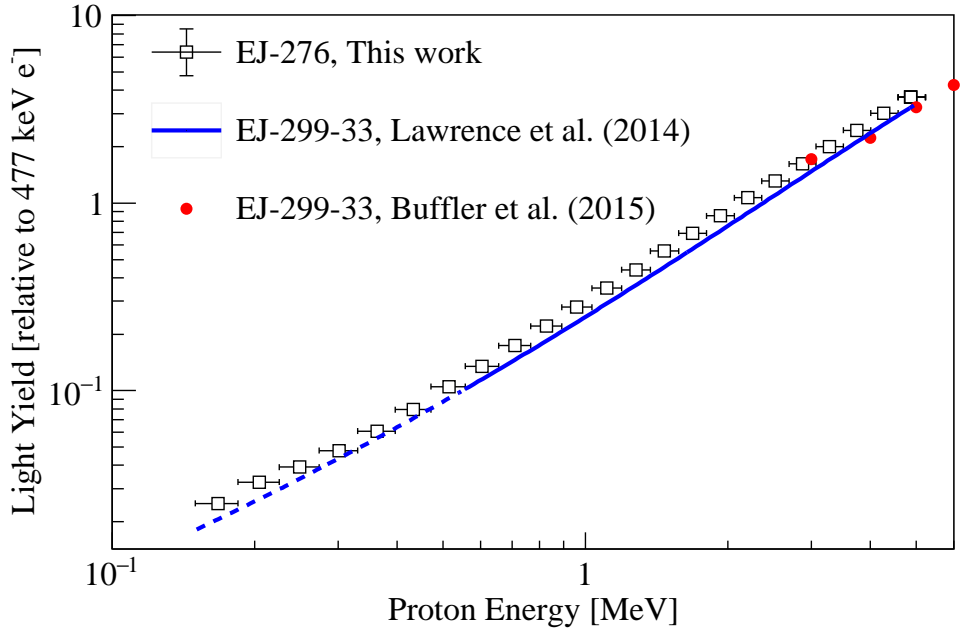}
	\caption{(Color online) EJ-276 proton light yield. For this work, the $x$-error bars represent a bin width and the $y$-error bars include both the statistical and systematic uncertainty. The functional form from Lawrence et al.\ (blue line) for EJ-299 is continuous over the measurement range and dashed in the extrapolation region.  \label{PLY276}}
\end{figure}

\begin{figure}
	\center
	\includegraphics[width=0.9\textwidth]{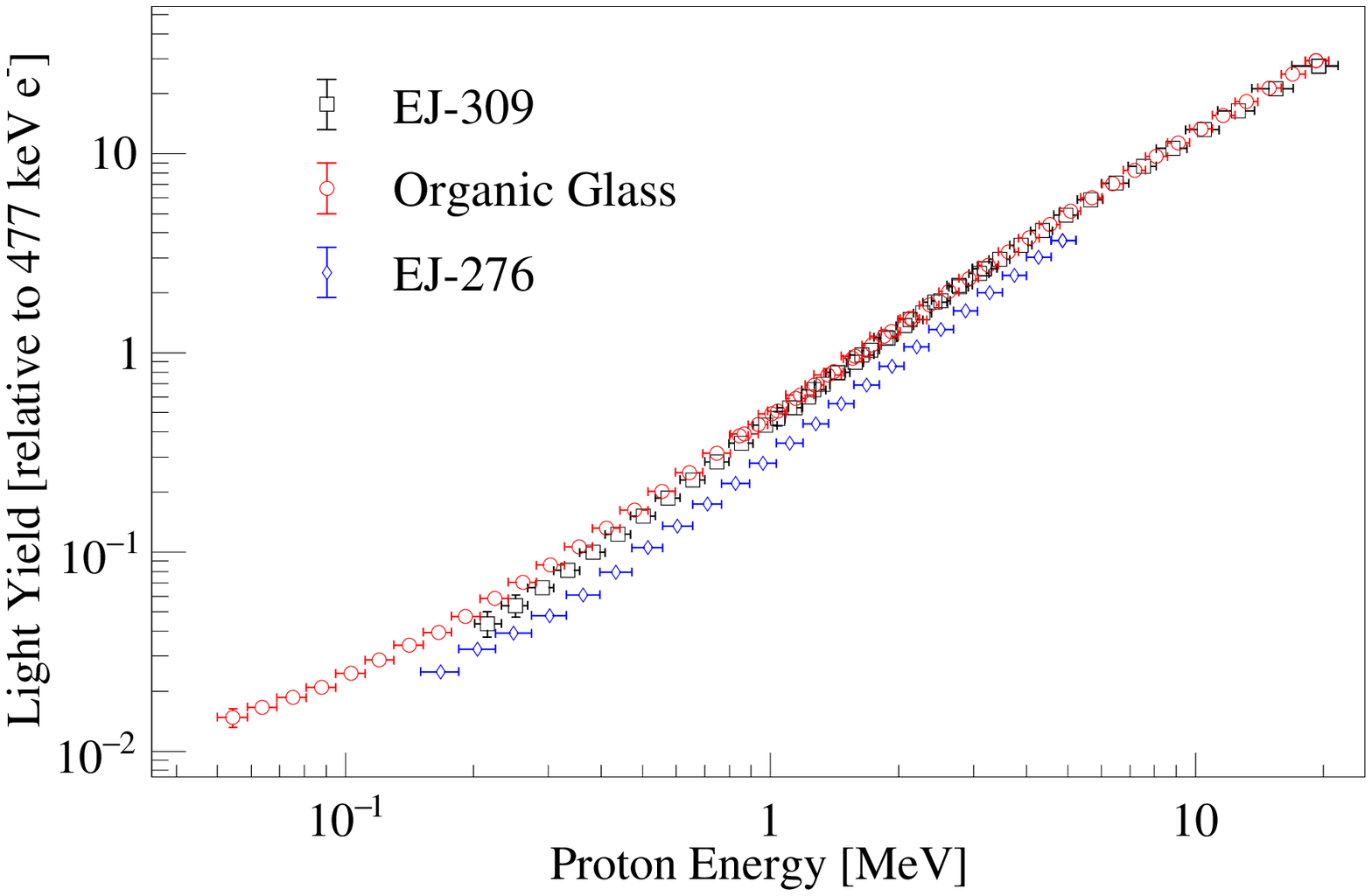}
	\caption{(Color online) Proton light yield of EJ-309, EJ-276, and the organic glass measured in this work. The $x$-error bars represent a bin width and the $y$-error bars include both the statistical and systematic uncertainty. \label{PLYAll}}
\end{figure}

Figure~\ref{PLY276} shows the EJ-276 proton light yield obtained in this work (open squares) from 167~keV to 4.9~MeV in comparison to its predecessor, EJ-299-33. This work shares a common energy range with the three lowest energy data points measured by Buffler et al.\ of the EJ-299-33 proton light yield using a 300~ns integration window (red circles)~\cite{Buffler2015_EJ299}. The lowest data point from Buffler et al.\ is in agreement with our measurement while the other two are approximately $15\%$ lower. Lawrence et al.\ measured the EJ-299-33 light yield for proton recoil energies ranging from 450 keV to 5.95 MeV. The empirical functional form fitted to the data provided by Lawrence et al.\ is plotted for the energy range of their measurement (continuous blue curve) and extrapolated over the range of the current measurement (dashed blue curve). The results have a similar shape to this measurement, but are approximately $11-18\%$ lower, likely due in part to the shorter integration length (180~ns, compared with the 350~ns integration length used in this work). 

Figure~\ref{PLYAll} shows a comparison of the proton light yield of EJ-276, EJ-309, and the organic glass obtained in this work. The measured data points and associated uncertainties are tabulated in appendix~\ref{LYDataPoints}. The relative proton light yield of EJ-309 and the organic glass are similar and within less than 10\% above 1 MeV. Below 1 MeV, the shape of the EJ-309 and organic glass proton light yield relations differs, and the organic glass exhibits a proton light yield higher than that of EJ-309. The proton light yield of EJ-276 is $25-35\%$ lower than that of EJ-309 over the energy range from 200 keV to 4 MeV. This significantly lower proton light yield should be taken into account when comparing the PSD performance of these materials when neutrons are the particle of interest. 


\section{Pulse shape discrimination properties}
\label{PSD_Measurement}

\begin{figure*}
	\centering
	\begin{subfigure}{0.75\textwidth}
		\centering
	   \includegraphics[height=6cm]{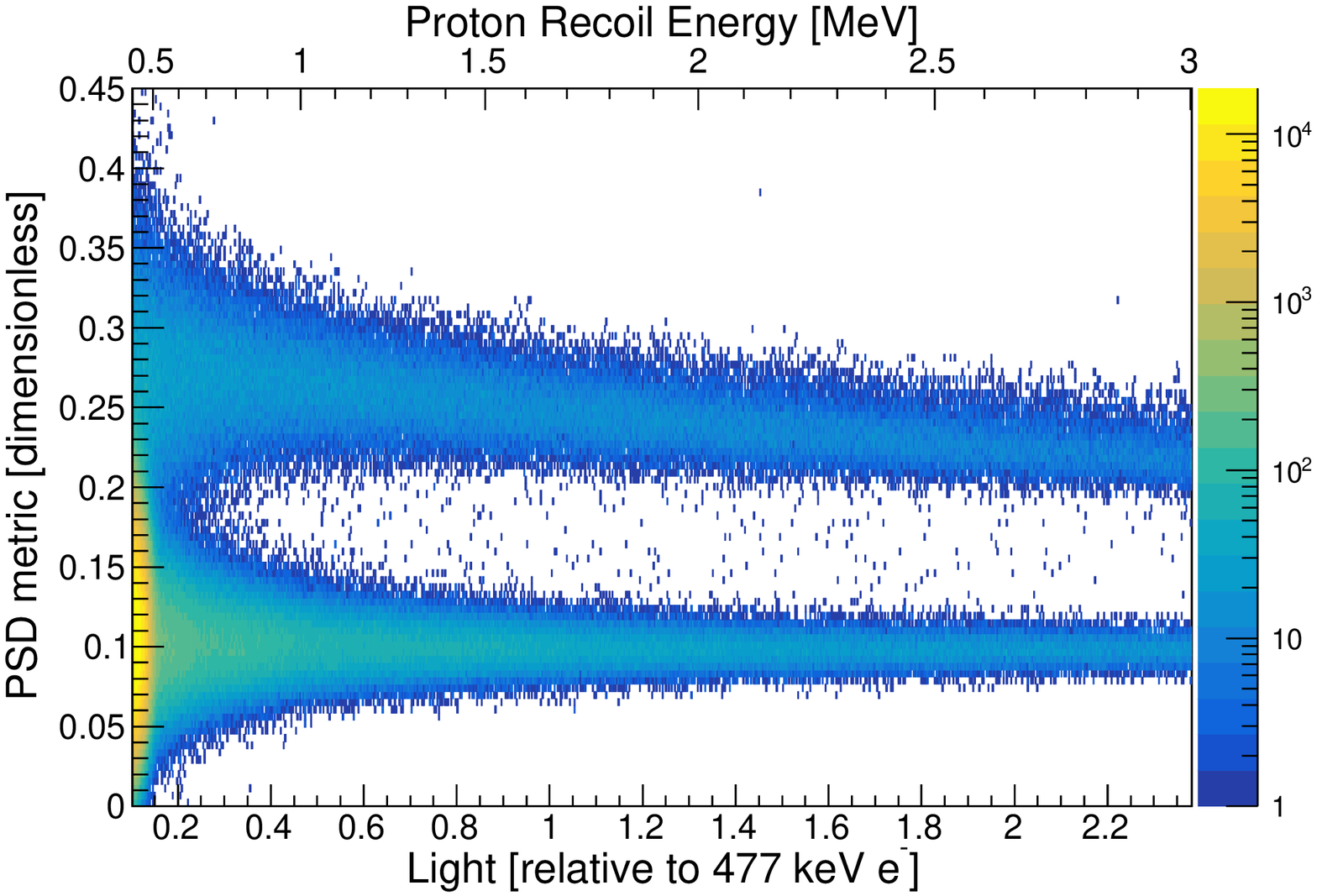} 
		\caption{EJ-309 with tail start at 38~ns and 270~ns total integration length.\label{lowest}}
	\end{subfigure}%
    \par
	\begin{subfigure}{0.75\textwidth}
		\centering
		\includegraphics[height=6cm]{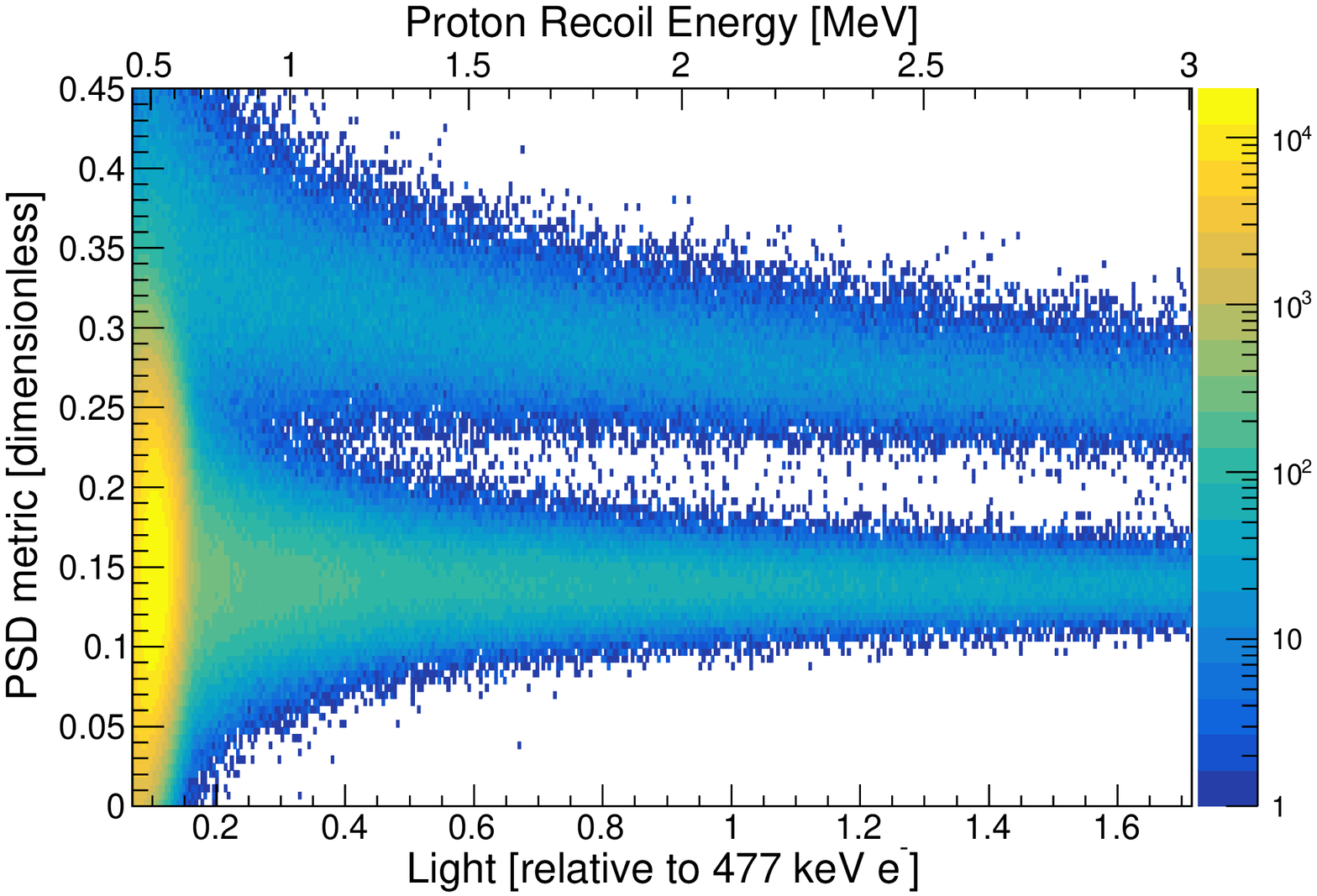}
		\caption{EJ-276 with tail start at 32~ns and 380~ns total integration length.}
	\end{subfigure}
    \par\bigskip
	\begin{subfigure}{0.75\textwidth}
		\centering
		\includegraphics[height=6cm]{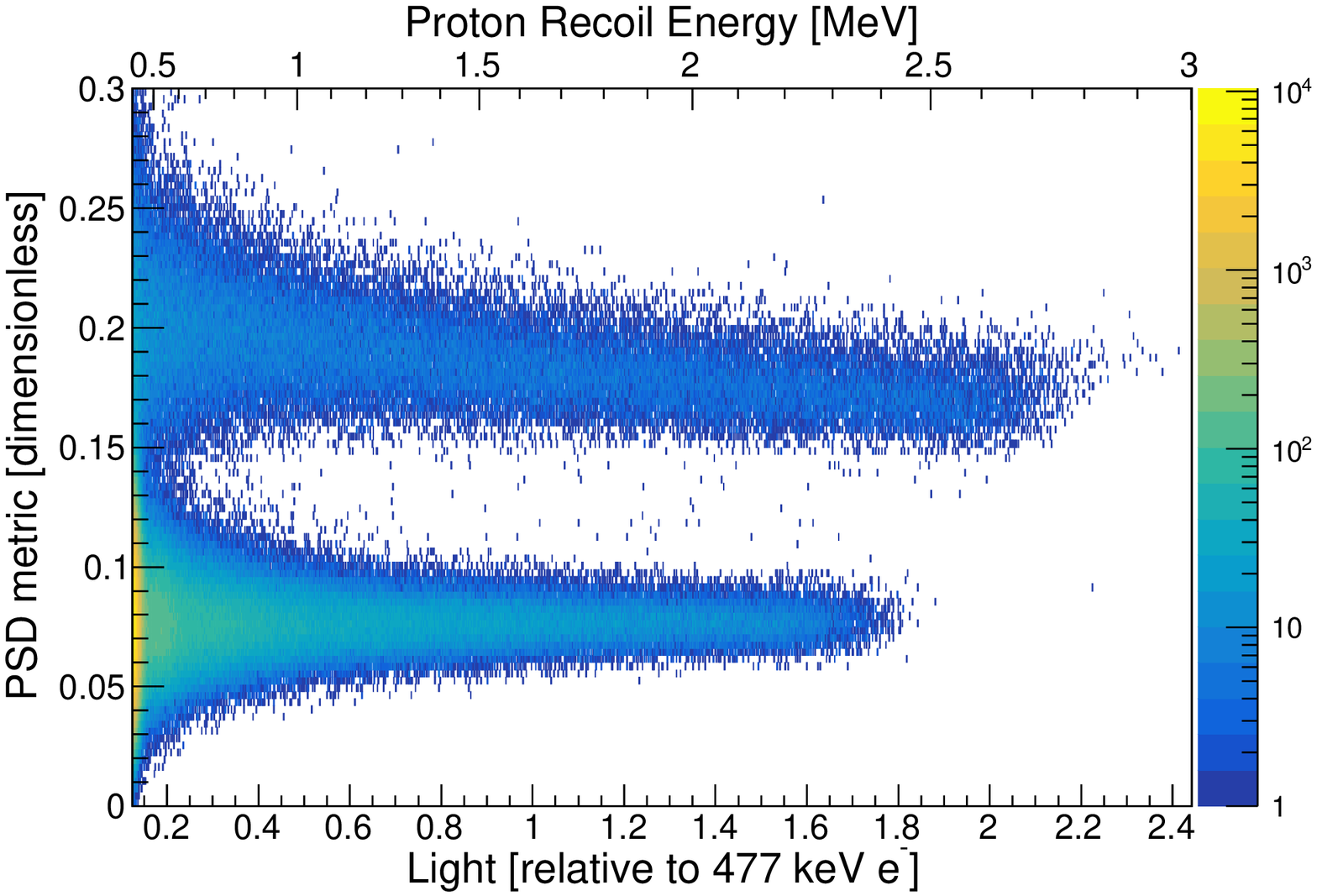}
		\caption{Organic glass with tail start at 30~ns and 240~ns total integration length.}
	\end{subfigure}
	\caption{(Color online) PSD metric, evaluated using tail-to-total, as function of the light output. The proton recoil energy, given on the secondary abscissa, is set to provide a common range for the three materials.
		\label{PSDplots}}
\end{figure*}

Neutron-$\gamma$ PSD performance is typically evaluated using a figure-of-merit (FOM) method. For example, using the charge integration approach, a PSD metric may be calculated on a per sample basis as the ratio of a tail integral to the integral of the full waveform~\cite{brooks1960}. The FOM is then given as~\cite{WinyardPSDFOM,Langeveld}:
\begin{linenomath*}
	\begin{equation}
	FOM = \frac{\mu_n-\mu_\gamma}{W_n+W_\gamma},
	\end{equation}
\end{linenomath*}
where $\mu_n$ and $\mu_{\gamma}$ are the centroids and $W_n$ and $W_{\gamma}$ are the full width at half maximum (FWHM) of the PSD metric distributions resulting from neutron and $\gamma$-induced events, respectively. When the FOM of two materials is compared on an electron-equivalent energy scale, the relative PSD performance may be misrepresented due to variations in ionization quenching between materials. These variations manifest as differences in the proportionality of light generated by neutron and $\gamma$-ray interactions, resulting in PSD performance metrics compared at the same electron-equivalent light output being biased towards materials with a lower relative proton light yield. Additionally, large discrepancies in FOM values exist in the literature due to variations in experimental setup, including differences in material size and shape, analog-to-digital converter bit resolution, sampling rate of digital acquisition, and mixed-field source spectra, among other factors~\cite{zaitseva2018, SODERSTROM2008, Cieslak}. Despite these challenges, when used in tandem with the proton light yield relation and for measurements conducted under the same experimental conditions, the FOM provides a means for quantitatively comparing the PSD performance of organic scintillators.

The neutron-$\gamma$ PSD performance of EJ-276, EJ-309, and the organic glass was evaluated using a tail-to-total charge integration method. To minimize sources of bias in comparing the PSD performance of the three materials, all scintillators were evaluated using the same active volume and cylindrical shape, with the scintillator coupling and housing reflectivity as described in section~\ref{materials}. The same PMT (Hamamatsu H1949-51) at the same bias voltage ($-1700$~V) was used with the same power supply (CAEN R1470ETD), digitizer, and pulse processing algorithms. A $\sim$53~mCi AmBe source was used, where the distance between the AmBe source and the detector was approximately 80~cm in all cases to limit pulse pile-up. Data were acquired over a period of approximately 2~h per scintillator.

Figure~\ref{PSDplots} shows two-dimensional plots of the PSD metric as a function of light output for (a) EJ-309, (b) EJ-276, and (c) the organic glass. The pulse integration length and tail start time were sampled over the range of 200 to 450~ns in 5~ns increments and 20 to 50~ns in 2~ns increments, respectively, and chosen to optimize the FOM at 1~MeV proton recoil energy. The relative light unit used for electron light output was determined as described in section~\ref{LOCalibration}. To quantify the discrimination performance in terms of neutron response, the secondary abscissa gives the proton recoil energy, with a common range for the three materials. The different ranges in the lower abscissa for the three materials arise from different ionization quenching effects manifested as differences in their proton light yield relations. While both the EJ-309 and organic glass scintillators show good separation between $\gamma$-ray and neutron events down to approximately 600~keV proton recoils, the EJ-276 medium lacks clear separation below 1~MeV proton recoils. 

\begin{figure}
	\center
	\includegraphics[width=0.9\textwidth]{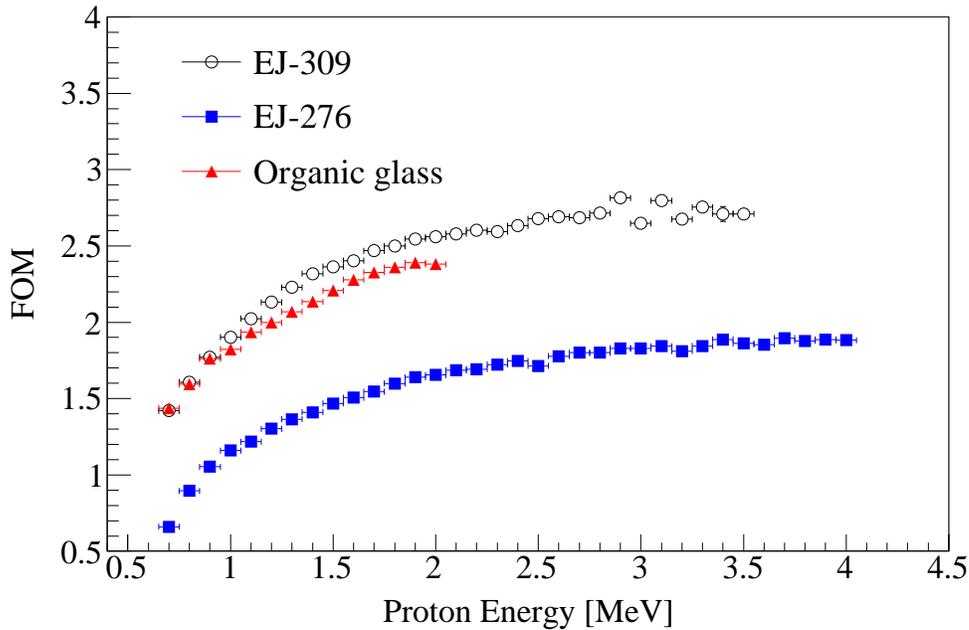}
	\caption{(Color online) FOM as a function of proton energy for EJ-309 (black), EJ-276 (blue), and the organic glass (red). The $x$-error bars reflect the 100-keV energy bin width. The $y$-error bars are smaller than the data points. Given a common supply voltage, a different proton energy range is covered for each of the three scintillators due to differences in their temporal response and proton light yield. \label{FOMResults}}
\end{figure}

Figure~\ref{FOMResults} shows the FOM as a function of proton recoil energy in 100 keV bins for the three materials.The error bars on the vertical axis are smaller than the data points. Given a common supply voltage, a different proton recoil energy range was covered for each of the three scintillators due to differences in their temporal response and proton light yield. For each material and in each bin, a parameter grid was established for the tail offset and integration length of the waveforms, and a grid search algorithm was used to optimize the FOM. The grid-search range for the pulse processing parameters varied for each material and as a function of energy. Histograms of the PSD metric were fit with two Gaussian distributions using a binned maximum likelihood algorithm with minimization performed using the ROOT Minuit package, and error matrices were obtained using the HESSE algorithm~\cite{brun}. The statistical uncertainty is less than $2\%$ and smaller than the data points. The PSD performance of the organic glass is superior to that of EJ-276, comparable to EJ-309 at low energies, and demonstrates good neutron-$\gamma$ separation at higher energies. The results obtained herein are in general agreement with recent measurements comparing the PSD performance of EJ-276 to that of EJ-301, a liquid organic scintillator from Eljen Technology with similar properties to EJ-309~\cite{grodzicka2020fast}. Previous measurements of a fresh EJ-276 sample by Zaitseva et al.\ found the PSD performance to be comparable to that of EJ-309~\cite{zaitseva2018}. The sample used in this work, however, had aged approximately one year at the time of measurement. As with the reduction in light output described in section~\ref{ElectronLO}, a similar degradation in PSD performance of the EJ-276 scintillator was observed.

\begin{figure}
	\center
	\includegraphics[width=0.9\textwidth]{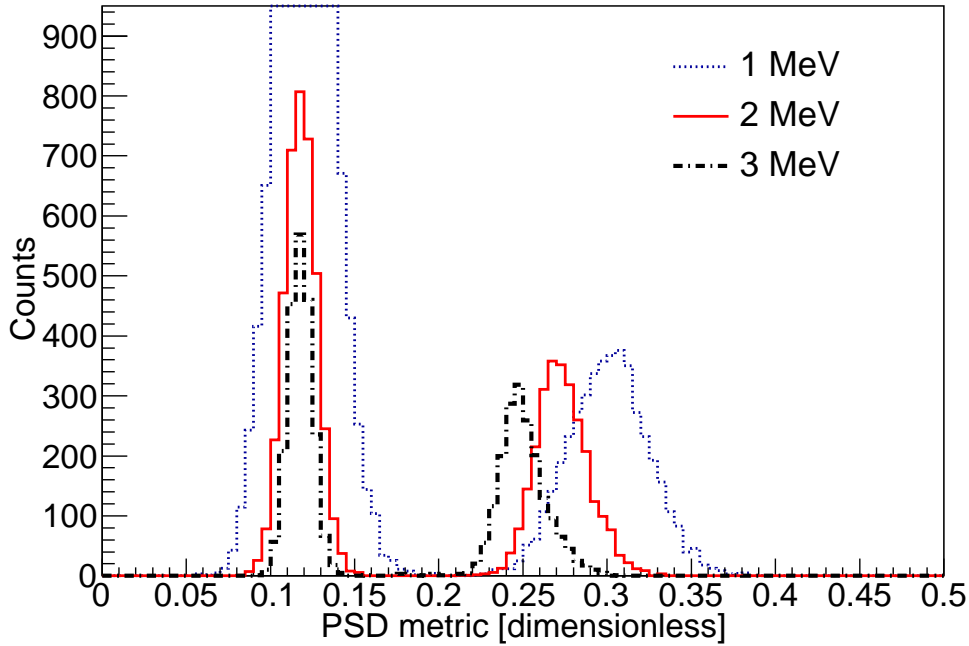}
	\caption{PSD metric distribution of EJ-309 for several different proton recoil energies. The centroid of the distribution of events resulting from proton recoils is energy dependent. \label{PSDShapeEvolution}}
\end{figure}

Figure~\ref{PSDShapeEvolution} shows the PSD metric distribution of EJ-309 for several different proton recoil energy bins of 100~keV width. For lower energies, both the neutron and $\gamma$-ray features broaden due to worsened resolution from lower photostatistics and degraded signal-to-noise. A variation in the centroid location of the neutron feature in the PSD metric distribution was observed as a function of energy, which reflects the energy-dependent temporal response of the scintillation light resulting from proton recoil interactions~\cite{Laplace-technote}. This results in a non-Gaussian skewing of the neutron feature as the energy bin width increases, which can result in a bias for FOMs determined using Gaussian fitting if the bin width is sufficiently large.

\section{Summary}
\label{conc}

This work presents measurements of the relative electron light output, proton light yield, and PSD performance of three organic scintillators: EJ-309, a liquid scintillator with good PSD properties and a low flash point, EJ-276, currently the only commercially available PSD-capable plastic scintillator, and an organic glass developed at Sandia National Laboratories. The relative electron light output was obtained by comparing the light output corresponding to the Compton edge of a 662~keV $\gamma$ ray. The organic glass demonstrated significantly higher electron light output in comparison to the other two materials; EJ-309 and EJ-276 were approximately $70\%$ and $50\%$ as bright as the organic glass, respectively. The proton light yield was measured using a double TOF technique over a broad energy range. The proton light yield of EJ-309 and the organic glass were within $10\%$ in the 1-20~MeV proton recoil energy range, whereas the proton light yield of EJ-276 was approximately $25-35\%$ lower than that of EJ-309 in the $0.2-4$~MeV energy range. Using the measured proton light yields, the PSD performance was quantified as a function of proton recoil energy using a FOM approach. The organic glass and EJ-309 demonstrated similar PSD performance as a function of proton recoil energy, with EJ-309 outperforming the organic glass for proton recoil energies above 1 MeV. Both showed significantly better discriminating power than EJ-276.

The organic glass developed at Sandia National Laboratories is a low cost, solid material with performance superior to EJ-276. In challenging environments, its physical characteristics may make it more suitable than the liquid scintillator, EJ-309. The higher luminosity of the organic glass can also allow for low energy recoil detection. This breakthrough material has the potential to replace common PSD-capable standards in a variety of fast neutron detection applications.

\appendix
\section{Proton light yield experimental details}
\label{appDetLoc}
The target and observation detector positions for the proton light yield measurements of EJ-309, EJ-276, and the organic glass (low and high bias voltage settings) are listed in tables~\ref{locationsDet309}, \ref{locationsDet276}, \ref{locationsOGlassLG}, and \ref{locationsOGlassHG}, respectively. A Cartesian coordinate system was used to reference detector locations (see figure~\ref{EJ276ExpSetup}) with the origin positioned at the beamline center on the west wall of the experimental area, which defines the $x=0$ plane. The $x$-coordinate was taken as the distance from the west wall, with the neutron production target located at beamline center with $x=-646.9~\pm~0.5$~cm. To determine detector locations within this coordinate system, a bi-plane laser was aligned with the beamline center. For the EJ-309 proton light yield measurement, the locations of all detectors were projected onto the floor using a plumb bob. A single plane laser in conjunction with a square was used to create a line from the detector location to the beamline center. The length along this line was taken as the $y$-coordinate of a given detector, and the distance along the beamline to the point of intersection was taken as the $x$-coordinate. For the EJ-276 and organic glass experimental campaigns, measurements of the $x$ and $y$ detector coordinates were made using a laser square along with a laser distance meter. All detector measurements were made to the center of the front face of the aluminum housing, whereas calculations of the flight path and angle were made using the center point of the scintillators. 

\begin{table*}[!h] 
	\caption{Summary of detector materials, identification (ID) numbers, locations, and estimated uncertainties in the measurements for the EJ-309 characterization. \label{locationsDet309}}
	\centering
	\begin{tabular}{ccccc}
		\hline
		Material    & ID &		x (cm)			&	y (cm)			&	z (cm)		\\ \hline
		EJ-309	& Target 		&		47.3	$\pm$	0.5& $	0.0	\pm	0.5	$   &$	\textcolor{white}{-}0.0	\pm	0.5	$\\ 
		EJ-309	& 0		&	125.7	$\pm$	1	& $	124.4	\pm	1	$&$	\textcolor{white}{-}0.0	\pm	0.2	$\\ 
		EJ-309	& 1 		&	151.7	$\pm$	1	&$	112.2	\pm	1	$&$	0.0	\pm	0.2	$\\ 
		EJ-309	& 2 		&	165.3 $\pm$	    1	&$	104.4	\pm	1	$&$	\textcolor{white}{-}4.0	\pm	0.2	$\\ 
		EJ-309	& 3 		&	165.3	$\pm$	1	&$	104.4	\pm	1	$&$	-1.5	\pm	0.2	$\\ 
		EJ-309	& 4		&	173.8	$\pm$	1	&$	92.1	\pm	1	$&$	\textcolor{white}{-}0.0	\pm	0.2	$\\ 
		EJ-309	& 5		&	184.0 $\pm$	    1	&$	77.5	\pm	1	$&$	0.0	\pm	0.2	$\\ 
		EJ-309	& 6 		&	203.6 $\pm$	    1	&$	60.6	\pm	1	$&$	\textcolor{white}{-}0.0	\pm	0.2	$\\ 
		\hline
	\end{tabular}
\end{table*}

\begin{table*}[!h]
	\caption{Summary of detector materials, ID numbers, locations, and estimated uncertainties in the measurements for the EJ-276 characterization. \label{locationsDet276}}
	\centering
	\begin{tabular}{ccccc}
		\hline
		Material    & ID &		x (cm)			&	y (cm)			&	z (cm)		\\ \hline
		EJ-276	& Target 		&		41.8	$\pm$	0.5& $	0.0	\pm	0.5	$   &$	\textcolor{white}{-}0.0	\pm	0.5	$\\ 
		EJ-309	& 0		&	146.0	$\pm$	1	& $	106.3	\pm	1	$&$	\textcolor{white}{-}10.9	\pm	0.2	$\\ 
		EJ-309	& 1 		&	153.8	$\pm$	1	&$	97.7	\pm	1	$&$	-11.1	\pm	0.2	$\\ 
		EJ-309	& 2 		&	165.4 $\pm$	    1	&$	87.2	\pm	1	$&$	\textcolor{white}{-}11.6	\pm	0.2	$\\ 
		EJ-309	& 3 		&	173.1	$\pm$	1	&$	78.3	\pm	1	$&$	-11.2	\pm	0.2	$\\ 
		EJ-309	& 4		&	186.3	$\pm$	1	&$	67.6	\pm	1	$&$	\textcolor{white}{-}9.7	\pm	0.2	$\\ 
		EJ-309	& 5		&	195.6 $\pm$	    1	&$	58.4	\pm	1	$&$	-9.2	\pm	0.2	$\\ 
		EJ-309	& 6 		&	205.0 $\pm$	    1	&$	50.0	\pm	1	$&$	\textcolor{white}{-}9.6	\pm	0.2	$\\ 
		EJ-309	& 7		&	223.0 $\pm$	    1	&$	31.4	\pm	1	$&$	-9.0	\pm	0.2	$\\ 
		EJ-309	& 8		&	224.7 $\pm$	    1	&$	30.7	\pm	1	$&$	\textcolor{white}{-}9.7	\pm	0.2	$\\ 
		\hline
	\end{tabular}
\end{table*}

\begin{table*}[!h] 
	\caption{Summary of detector materials, ID numbers, locations, and estimated uncertainties in the measurements of the organic glass at low gain settings. \label{locationsOGlassLG}}
	\centering
	\begin{tabular}{ccccc}
		\hline
		Material    & ID &		x (cm)			&	y (cm)			&	z (cm)		\\ \hline
		Organic Glass	& Target 		&		67.3	$\pm$	0.5& $	0.0	\pm	0.5	$   &$	\textcolor{white}{-}0.3	\pm	0.5	$\\ 
		EJ-309	& 0		&	108.7	$\pm$	1	& $	129.9	\pm	1	$&$	\textcolor{white}{-}15.2	\pm	0.2	$\\ 
		EJ-309	& 1 		&	139.8	$\pm$	1	&$	117.0	\pm	1	$&$	\textcolor{white}{-} 14.8	\pm	0.2	$\\ 
		EJ-309	& 2 		&	160.0 $\pm$	    1	&$	109.0	\pm	1	$&$	\textcolor{white}{-}14.8	\pm	0.2	$\\ 
		EJ-309	& 3 		&	181.0	$\pm$	1	&$	99.6	\pm	1	$&$	\textcolor{white}{-}14.6	\pm	0.2	$\\ 
		EJ-309	& 4		&	203.6	$\pm$	1	&$	92.1	\pm	1	$&$	\textcolor{white}{-}12.8	\pm	0.2	$\\ 
		EJ-309	& 5		&	226.1 $\pm$	    1	&$	82.7	\pm	1	$&$	\textcolor{white}{-}13.1	\pm	0.2	$\\ 
		EJ-309	& 6 		&	135.7 $\pm$	    1	&$	120.5	\pm	1	$&$	-5.8	\pm	0.2	$\\ 
		EJ-309	& 7		&	152.5 $\pm$	    1	&$	113.3	\pm	1	$&$	-5.9	\pm	0.2	$\\ 
		EJ-309	& 8		&	173.1 $\pm$	    1	&$	104.6	\pm	1	$&$	-6.0	\pm	0.2	$\\ 
		EJ-309	& 9		&	194.6 $\pm$	    1	&$	95.8	\pm	1	$&$	-6.0	\pm	0.2	$\\ 
		EJ-309	& 10		&	213.8 $\pm$	    1	&$	88.2	\pm	1	$&$	-5.8	\pm	0.2	$\\ 
		\hline
	\end{tabular}
\end{table*}

\begin{table*}[!h]
	\caption{Summary of detector materials, ID numbers, locations, and estimated uncertainties in the measurements of the organic glass at high gain settings. \label{locationsOGlassHG}}
	\centering
	\begin{tabular}{ccccc}
		\hline
		Material    & ID &		x (cm)			&	y (cm)			&	z (cm)		\\ \hline
		Organic Glass	& Target 		&		59.1	$\pm$	0.5& $	0.0	\pm	0.5	$   &$	\textcolor{white}{-}0.3	\pm	0.5	$\\ 
		EJ-309	& 0		&	174.6	$\pm$	1	& $	91.7	\pm	1	$&$	\textcolor{white}{-}14.9	\pm	0.2	$\\ 
		EJ-309	& 1 		&	197.0	$\pm$	1	&$	82.5	\pm	1	$&$	\textcolor{white}{-} 15.1	\pm	0.2	$\\ 
		EJ-309	& 2 		&	217.8 $\pm$	    1	&$	74.1	\pm	1	$&$	\textcolor{white}{-}14.9	\pm	0.2	$\\ 
		EJ-309	& 3 		&	237.3	$\pm$	1	&$	66.2	\pm	1	$&$	\textcolor{white}{-}14.6	\pm	0.2	$\\ 
		EJ-309	& 4		&	263.8	$\pm$	1	&$	57.1	\pm	1	$&$	\textcolor{white}{-}13.0	\pm	0.2	$\\ 
		EJ-309	& 5		&	289.0 $\pm$	    1	&$	46.7	\pm	1	$&$	\textcolor{white}{-}12.7	\pm	0.2	$\\ 
		EJ-309	& 6		&	188.9 $\pm$	   1	&$	87.5	\pm	1	$&$	-5.8	\pm	0.2	$\\ 
		EJ-309	& 7		&	213.1 $\pm$	    1	&$	77.5	\pm	1	$&$	-5.8	\pm	0.2	$\\ 
		EJ-309	& 8		&	236.4 $\pm$	    1	&$	68.2	\pm	1	$&$	-5.7	\pm	0.2	$\\ 
		EJ-309	& 9		&	261.9 $\pm$	    1	&$	58.1	\pm	1	$&$	-5.6	\pm	0.2	$\\ 
		EJ-309	& 10 		&	292.5 $\pm$	    1	&$	45.4	\pm	1	$&$	-6.1	\pm	0.2	$\\ 
		
		\hline
	\end{tabular}
\end{table*}

Table~\ref{timeCalibData} provides the standard deviation of the fitted normal distributions of the time difference histograms used for calibration of the incoming and outgoing TOF (as described in section~\ref{TOFCalib}).

\begin{table*} 
	\caption{Summary of the uncertainties in the incoming and outgoing TOF calibrations, where $\sigma_{\mathrm{IncTOF}}$ corresponds to the standard deviation of the fitted normal distribution to the peak feature in the historgram of time differences between the cyclotron RF signal and $\gamma$-ray events in the target detector (e.g., see figure~\ref{incTOF}) and $\sigma_{\mathrm{OutTOF}}$ corresponds to the standard deviation of the fitted normal distribution to the peak feature in the histogram of time differences between $\gamma$-ray events in the target and observation detectors. \label{timeCalibData}}
	\centering
	\begin{tabular}{cccc}
		\hline
		& $\sigma_{\mathrm{IncTOF}}$	[ns] & $\min$ ($\sigma_{\mathrm{OutTOF}}$) [ns] & $\max$ ($\sigma_{\mathrm{OutTOF}}$) [ns]\\
		\hline
		EJ-276 & 3.28 & 0.37 & 0.48 \\
		EJ-309 & 3.78 & 0.43 & 0.56 \\
		Organic glass (High PMT gains) & 2.24 & 0.33 & 0.46 \\
		Organic glass (Low PMT gains) & 3.15 & 0.38 & 0.46 \\
		\hline
	\end{tabular}
\end{table*}

\section{Proton light yield data}
\label{LYDataPoints}

The proton light yield data for EJ-309 and EJ-276 are summarized in table~\ref{PLYResultsTable}. The organic glass proton light yield data for the high and low bias voltage settings are summarized in table~\ref{PLYResultsTable2}. The asymmetric proton recoil energy bin widths are reflective of the non-uniform distribution of proton energies within a given bin.

\begin{table*}[!h]
	\caption{Relative proton light yield data for EJ-276 and EJ-309. Proton recoil energy bin widths are provided as well as the light output uncertainties. Covariance matrices are available upon request.}
\label{PLYResultsTable}
	\centering
	\setlength{\tabcolsep}{8pt}
	\begin{tabular}{cc|cc}
		\hline
		\multicolumn{2}{c|}{EJ-276} & \multicolumn{2}{c}{EJ-309}\\
		Proton Recoil   &  Light Yield  & Proton Recoil   &  Light Yield   \\
		Energy [MeV] &  [rel. 477 keV electron]   & Energy [MeV] &  [rel. 477 keV electron] \\
\hline
0.167$_{-0.017}^{+0.017}$ & 0.0250 $\pm$ 0.0014 & 0.216$_{-0.014}^{+0.017}$ & 0.0437 $\pm$ 0.0063 \\ 
0.205$_{-0.020}^{+0.021}$ & 0.0326 $\pm$ 0.0015 & 0.252$_{-0.019}^{+0.017}$ & 0.0538 $\pm$ 0.0068 \\ 
0.249$_{-0.024}^{+0.025}$ & 0.0391 $\pm$ 0.0016 & 0.290$_{-0.022}^{+0.019}$ & 0.0660 $\pm$ 0.0039 \\ 
0.302$_{-0.028}^{+0.029}$ & 0.0479 $\pm$ 0.0020 & 0.333$_{-0.024}^{+0.023}$ & 0.0809 $\pm$ 0.0039 \\ 
0.363$_{-0.032}^{+0.033}$ & 0.0608 $\pm$ 0.0024 & 0.382$_{-0.027}^{+0.026}$ & 0.0997 $\pm$ 0.0048 \\ 
0.433$_{-0.037}^{+0.039}$ & 0.0792 $\pm$ 0.0037 & 0.438$_{-0.030}^{+0.030}$ & 0.1226 $\pm$ 0.0058 \\ 
0.515$_{-0.043}^{+0.043}$ & 0.1051 $\pm$ 0.0043 & 0.502$_{-0.034}^{+0.033}$ & 0.1515 $\pm$ 0.0065 \\ 
0.604$_{-0.047}^{+0.051}$ & 0.1349 $\pm$ 0.0053 & 0.573$_{-0.038}^{+0.039}$ & 0.1866 $\pm$ 0.0083 \\ 
0.710$_{-0.055}^{+0.056}$ & 0.1737 $\pm$ 0.0068 & 0.656$_{-0.044}^{+0.044}$ & 0.2303 $\pm$ 0.0105 \\ 
0.827$_{-0.061}^{+0.064}$ & 0.2206 $\pm$ 0.0079 & 0.748$_{-0.049}^{+0.050}$ & 0.2840 $\pm$ 0.0133 \\ 
0.959$_{-0.067}^{+0.074}$ & 0.2785 $\pm$ 0.0103 & 0.853$_{-0.055}^{+0.057}$ & 0.3505 $\pm$ 0.0159 \\ 
1.110$_{-0.078}^{+0.082}$ & 0.3518 $\pm$ 0.0123 & 0.974$_{-0.064}^{+0.063}$ & 0.4323 $\pm$ 0.0205 \\ 
1.278$_{-0.087}^{+0.093}$ & 0.4414 $\pm$ 0.0150 & 1.109$_{-0.071}^{+0.074}$ & 0.5290 $\pm$ 0.0258 \\ 
1.469$_{-0.098}^{+0.104}$ & 0.5548 $\pm$ 0.0187 & 1.263$_{-0.081}^{+0.084}$ & 0.6511 $\pm$ 0.0330 \\ 
1.684$_{-0.110}^{+0.119}$ & 0.6912 $\pm$ 0.0224 & 1.439$_{-0.092}^{+0.095}$ & 0.7973 $\pm$ 0.0407 \\ 
1.928$_{-0.126}^{+0.134}$ & 0.8554 $\pm$ 0.0279 & 1.639$_{-0.104}^{+0.111}$ & 0.9789 $\pm$ 0.0512 \\ 
2.204$_{-0.142}^{+0.152}$ & 1.0671 $\pm$ 0.0343 & 1.867$_{-0.118}^{+0.128}$ & 1.1938 $\pm$ 0.0641 \\ 
2.518$_{-0.159}^{+0.175}$ & 1.3125 $\pm$ 0.0406 & 2.131$_{-0.135}^{+0.147}$ & 1.4641 $\pm$ 0.0815 \\ 
2.873$_{-0.182}^{+0.200}$ & 1.6199 $\pm$ 0.0511 & 2.434$_{-0.156}^{+0.169}$ & 1.7903 $\pm$ 0.1025 \\ 
3.278$_{-0.206}^{+0.229}$ & 2.0029 $\pm$ 0.0632 & 2.779$_{-0.176}^{+0.199}$ & 2.1857 $\pm$ 0.1291 \\ 
3.740$_{-0.233}^{+0.265}$ & 2.4430 $\pm$ 0.0736 & 3.180$_{-0.202}^{+0.232}$ & 2.6460 $\pm$ 0.1411 \\ 
4.268$_{-0.262}^{+0.308}$ & 3.0231 $\pm$ 0.0972 & & \\ 
4.869$_{-0.293}^{+0.362}$ & 3.6697 $\pm$ 0.1119 & & \\ 
		\hline
	\end{tabular}
\end{table*}

\begin{table*}[!h]
	\caption{Relative proton light yield data for the organic glass. Proton recoil energy bin widths are provided as well as the light output uncertainties. Covariance matrices are available upon request.}
\label{PLYResultsTable2}
	\centering
	\setlength{\tabcolsep}{8pt}
	\begin{tabular}{cc|cc}
		\hline
		\multicolumn{2}{c|}{High PMT gains} & \multicolumn{2}{c}{Low PMT gains}\\
		Proton Recoil   & Light Yield  & Proton Recoil   & Light Yield   \\
		Energy [MeV] &  [rel. 477 keV electron]   & Energy [MeV] &  [rel. 477 keV electron] \\
		\hline
		0.054$_{-0.004}^{+0.004}$ & 0.0148 $\pm$ 0.0016 & 0.844$_{-0.044}^{+0.043}$ & 0.3841$\pm$0.0097 \\       
		0.064$_{-0.005}^{+0.005}$ & 0.0166 $\pm$ 0.0009 & 0.935$_{-0.048}^{+0.048}$ & 0.4355$\pm$0.0113 \\       
		0.075$_{-0.006}^{+0.006}$ & 0.0186 $\pm$ 0.0011 & 1.038$_{-0.055}^{+0.051}$ & 0.5097$\pm$0.0135 \\       
		0.088$_{-0.007}^{+0.007}$ & 0.0209 $\pm$ 0.0007 & 1.147$_{-0.058}^{+0.059}$ & 0.5896$\pm$0.0142 \\       
		0.103$_{-0.008}^{+0.008}$ & 0.0246 $\pm$ 0.0014 & 1.270$_{-0.063}^{+0.067}$ & 0.6888$\pm$0.0159 \\       
		0.120$_{-0.009}^{+0.010}$ & 0.0287 $\pm$ 0.0013 & 1.409$_{-0.072}^{+0.072}$ & 0.8030$\pm$0.0191 \\       
		0.141$_{-0.011}^{+0.011}$ & 0.0341 $\pm$ 0.0013 & 1.560$_{-0.079}^{+0.081}$ & 0.9370$\pm$0.0206 \\       
		0.166$_{-0.013}^{+0.012}$ & 0.0395 $\pm$ 0.0014 & 1.729$_{-0.089}^{+0.090}$ & 1.092$\pm$0.025\\       
		0.192$_{-0.014}^{+0.016}$ & 0.0475 $\pm$ 0.0017 & 1.922$_{-0.103}^{+0.096}$ & 1.282$\pm$0.027\\       
		0.224$_{-0.017}^{+0.017}$ & 0.0584 $\pm$ 0.0022 & 2.129$_{-0.111}^{+0.111}$ & 1.489$\pm$0.033\\       
		0.262$_{-0.020}^{+0.020}$ & 0.0706 $\pm$ 0.0024 & 2.365$_{-0.124}^{+0.125}$ & 1.735$\pm$0.036\\       
		0.304$_{-0.022}^{+0.024}$ & 0.0860 $\pm$ 0.0028 & 2.630$_{-0.140}^{+0.139}$ & 2.029$\pm$0.041\\       
		0.354$_{-0.027}^{+0.027}$ & 0.1062 $\pm$ 0.0034 & 2.925$_{-0.155}^{+0.159}$ & 2.362$\pm$0.047\\       
		0.411$_{-0.030}^{+0.031}$ & 0.1315 $\pm$ 0.0040 & 3.258$_{-0.175}^{+0.178}$ & 2.758$\pm$0.052\\       
		0.479$_{-0.037}^{+0.035}$ & 0.1622 $\pm$ 0.0050 & 3.631$_{-0.195}^{+0.203}$ & 3.217$\pm$0.059\\       
		0.555$_{-0.041}^{+0.042}$ & 0.2008 $\pm$ 0.0062 & 4.057$_{-0.222}^{+0.228}$ & 3.768$\pm$0.068\\       
		0.644$_{-0.047}^{+0.049}$ & 0.2510 $\pm$ 0.0076 & 4.535$_{-0.250}^{+0.259}$ & 4.389$\pm$0.075\\       
		0.747$_{-0.054}^{+0.057}$ & 0.3126 $\pm$ 0.0092 & 5.082$_{-0.288}^{+0.289}$ & 5.143$\pm$0.087\\       
		0.867$_{-0.063}^{+0.067}$ & 0.3907 $\pm$ 0.0110 & 5.695$_{-0.324}^{+0.330}$ & 6.021$\pm$0.099\\       
		1.008$_{-0.074}^{+0.078}$ & 0.4924 $\pm$ 0.0144 & 6.389$_{-0.364}^{+0.380}$ & 7.047$\pm$0.111\\       
		1.174$_{-0.088}^{+0.088}$ & 0.6138 $\pm$ 0.0185 & 7.184$_{-0.415}^{+0.432}$ & 8.252$\pm$0.126\\       
		1.366$_{-0.103}^{+0.103}$ & 0.7757 $\pm$ 0.0219 & 8.090$_{-0.474}^{+0.491}$ & 9.681$\pm$0.147\\       
		1.585$_{-0.116}^{+0.126}$ & 0.9611 $\pm$ 0.0263 & 9.113$_{-0.533}^{+0.569}$ & 11.32$\pm$0.17 \\       
		1.848$_{-0.136}^{+0.148}$ & 1.2129 $\pm$ 0.0321 & 10.29$_{-0.61}^{+0.65}$ & 13.30$\pm$0.19 \\       
		2.155$_{-0.160}^{+0.175}$ & 1.4682 $\pm$ 0.0309 & 11.61$_{-0.67}^{+0.77}$ & 15.55$\pm$0.21 \\       
		&                     & 13.17$_{-0.79}^{+0.86}$ & 18.25$\pm$0.25 \\
		&                     & 14.90$_{-0.87}^{+1.02}$ & 21.36$\pm$0.29 \\
		&                     & 16.90$_{-0.98}^{+1.19}$ & 25.00$\pm$0.34 \\
		&                     & 19.22$_{-1.13}^{+1.37}$ & 29.16$\pm$0.41 \\
		\hline
	\end{tabular}
\end{table*}
\clearpage

\acknowledgments
The authors thank the 88-Inch Cyclotron operations and facilities staff for their help in performing these experiments and gratefully acknowledge Dr. Natalia Zaitseva for valuable discussions. This work was performed under the auspices of the U.S. Department of Energy National Nuclear Security Administration by Lawrence Berkeley National Laboratory under Contract DE-AC02-05CH11231 and through the Nuclear Science and Security Consortium under Award No. DE-NA0003180. Sandia National Laboratories is a multimission laboratory managed and operated by National Technology and Engineering Solutions of Sandia LLC, a wholly owned subsidiary of Honeywell International Inc., for the U.S. Department of Energy's National Nuclear Security Administration under contract DE-NA0003525.	

\bibliographystyle{unsrt}
\bibliography{PSDScintCharacterization}

\end{document}